\newif\iffigs\figstrue
\documentclass[12pt,epsfig]{article}
\iffigs
 \usepackage{epsfig}
\else
\fi
\newcommand{\eqn}[1]{(\ref{#1})}

\newsavebox{\uuunit}
\sbox{\uuunit}
    {\setlength{\unitlength}{0.825em}
     \begin{picture}(0.6,0.7)
        \thinlines
        \put(0,0){\line(1,0){0.5}}
        \put(0.15,0){\line(0,1){0.7}}
        \put(0.35,0){\line(0,1){0.8}}
       \multiput(0.3,0.8)(-0.04,-0.02){12}{\rule{0.5pt}{0.5pt}}
     \end {picture}}

\def\Im{{\rm Im ~}}
\def\Re{{\rm Re ~}}
\def\IP{\relax{\rm I\kern-.18em P}}

\begin{document}
%
\font\cmss=cmss10 \font\cmsss=cmss10 at 7pt
\def\twomat#1#2#3#4{\left(\matrix{#1 & #2 \cr #3 & #4}\right)}
\def\inbar{\vrule height1.5ex width.4pt depth0pt}
\def\IC{{\rm C}}
\def\IG{\relax\,\hbox{$\inbar\kern-.3em{\rm G}$}}
\def\IB{\relax{\rm I\kern-.18em B}}
\def\ID{\relax{\rm I\kern-.18em D}}
\def\IL{\relax{\rm I\kern-.18em L}}
\def\IF{\relax{\rm I\kern-.18em F}}
\def\IH{\relax{\rm I\kern-.18em H}}
\def\II{\relax{\rm I\kern-.17em I}}
\def\IN{\relax{\rm I\kern-.18em N}}
\def\IP{\relax{\rm I\kern-.18em P}}
\def\IQ{\relax\,\hbox{$\inbar\kern-.3em{\rm Q}$}}
\def\bfzero{\relax\,\hbox{$\inbar\kern-.3em{\rm 0}$}}
\def\IK{\relax{\rm I\kern-.18em K}}
\def\IG{\relax\,\hbox{$\inbar\kern-.3em{\rm G}$}}
 \font\cmss=cmss10 \font\cmsss=cmss10 at 7pt
\def\IR{\relax{\rm I\kern-.18em R}}
\def\ZZ{\relax\ifmmode\mathchoice
{\hbox{\cmss Z\kern-.4em Z}}{\hbox{\cmss Z\kern-.4em Z}}
{\lower.9pt\hbox{\cmsss Z\kern-.4em Z}}
{\lower1.2pt\hbox{\cmsss Z\kern-.4em Z}}\else{\cmss Z\kern-.4em
Z}\fi}
\def\bfone{\relax{\rm 1\kern-.35em 1}}
\def\dop{{\rm d}\hskip -1pt}
\def\real{{\rm Re}\hskip 1pt}
\def\trace{{\rm Tr}\hskip 1pt}
\def\ii{{\rm i}}
\def\diag{{\rm diag}}
\def\sch#1#2{\{#1;#2\}}
\def\bfone{\relax{\rm 1\kern-.35em 1}}
\font\cmss=cmss10 \font\cmsss=cmss10 at 7pt
\def\a{\alpha} \def\b{\beta} \def\d{\delta}
\def\e{\epsilon} \def\c{\gamma}
\def\G{\Gamma} \def\l{\lambda}
\def\L{\Lambda} \def\s{\sigma}
\def\cA{{\cal A}} \def\cB{{\cal B}}
\def\cC{{\cal C}} \def\cD{{\cal D}}
\def\cF{{\cal F}} \def\cG{{\cal G}}
\def\cH{{\cal H}} \def\cI{{\cal I}}
\def\cJ{{\cal J}} \def\cK{{\cal K}}
\def\cL{{\cal L}} \def\cM{{\cal M}}
\def\cN{{\cal N}} \def\cO{{\cal O}}
\def\cP{{\cal P}} \def\cQ{{\cal Q}}
\def\cR{{\cal R}} \def\cV{{\cal V}}\def\cW{{\cal W}}
\newcommand{\be}{\begin{equation}}
\newcommand{\ee}{\end{equation}}
\newcommand{\bea}{\begin{eqnarray}}
\newcommand{\eea}{\end{eqnarray}}
\let\la=\label \let\ci=\cite \let\re=\ref
%
%
%
\def\crr{\crcr\noalign{\vskip {8.3333pt}}}
\def\tilde{\widetilde}
\def\bar{\overline}
\def\us#1{\underline{#1}}
\let\shat=\hat
\def\hat{\widehat}
\def\hyp{\vrule height 2.3pt width 2.5pt depth -1.5pt}
\def\square{\mbox{.08}{.08}}
\def\Coeff#1#2{{#1\over #2}}
\def\Coe#1.#2.{{#1\over #2}}
\def\coeff#1#2{\relax{\textstyle {#1 \over #2}}\displaystyle}
\def\coe#1.#2.{\relax{\textstyle {#1 \over #2}}\displaystyle}
\def\half{{1 \over 2}}
\def\shalf{\relax{\textstyle {1 \over 2}}\displaystyle}
\def\dag#1{#1\!\!\!/\,\,\,}
\def\to{\rightarrow}
\def\notin{\hbox{{$\in$}\kern-.51em\hbox{/}}}
\def\shdot{\!\cdot\!}
\def\ket#1{\,\big|\,#1\,\big>\,}
\def\bra#1{\,\big<\,#1\,\big|\,}
\def\equaltop#1{\mathrel{\mathop=^{#1}}}
\def\Trbel#1{\mathop{{\rm Tr}}_{#1}}
\def\inserteq#1{\noalign{\vskip-.2truecm\hbox{#1\hfil}
\vskip-.2cm}}
\def\attac#1{\Bigl\vert
{\phantom{X}\atop{{\rm\scriptstyle #1}}\phantom{X}}}
\def\exx#1{e^{{\displaystyle #1}}}
\def\del{\partial}
\def\delbar{\bar\partial}
\def\nex#1{$N\!=\!#1$}
\def\dex#1{$d\!=\!#1$}
\def\cex#1{$c\!=\!#1$}
\def\eg{{\it e.g.}} \def\ie{{\it i.e.}}
\def\IE{\relax{{\rm I\kern-.18em E}}}
\def\cE{{\cal E}}
\def\rt{{\cR^{(3)}}}
\def\IGam{\relax{{\rm I}\kern-.18em \Gamma}}
\def\IGa{\IA}
\def\ii{{\rm i}}
\begin{titlepage}
\begin{center}
{\LARGE $N=8$ BPS Black Holes with  $1/2$ or $1/4$ supersymmetry\\
\vspace{6pt}
and\\
\vspace{6pt}
Solvable Lie algebra decompositions
 $^*$ }\\
\vfill
{\large G. Arcioni$^1$, A. Ceresole$^2$, F. Cordaro$^1$,
\\
\vskip 0.2cm
R. D'Auria $^2$, P. Fr\'e$^1$, L. Gualtieri$^1$
 and  M. Trigiante$^3$   } \\
\vfill
{\small
$^1$ Dipartimento di Fisica Teorica, Universit\'a di Torino, via P. Giuria 1,
I-10125 Torino, \\
Istituto Nazionale di Fisica Nucleare (INFN) - Sezione di Torino, Italy \\
\vspace{6pt}
$^2$ Dipartimento di Fisica, Politecnico di Torino, C.so Duca degli
Abruzzi, 24, I-10129 Torino\\
and Istituto Nazionale di Fisica Nucleare (INFN) - Sezione di Torino, Italy\\
\vspace{6pt}
$^3$ Department of Physics, University of Wales Swansea, Singleton Park,\\
Swansea SA2 8PP, United Kingdom\\
\vspace{6pt}
}
\end{center}
\vfill
\begin{center}
{\bf Abstract}
\end{center}
{\small In the context of $N=8$ supergravity we construct the general form of
BPS 0--branes that preserve either $1/2$ or $1/4$ of the original supersymmetry.
We show how such solutions are related to suitable decompositions of
the $70$ dimensional solvable Lie algebra that describes the scalar field
sector. We compare our new results to those obtained in a previous
paper for the case of $1/8$ supersymmetry preserving black holes.
Each of the three cases is based on a different solvable Lie algebra
decomposition and leads to a different structure of the scalar field
evolution and of their fixed  values at the horizon of the black
hole.
}
\vspace{2mm} \vfill \hrule width 3.cm
{\footnotesize
 $^*$ Supported in part by   EEC  under TMR contract
 ERBFMRX-CT96-0045, in which R.D'Auria and A. Ceresole are associated to Torino University
 and by TMR contract ERBMRXCT960012 in which M. Trigiante is associated
 to Swansea University}
\end{titlepage}
\eject
\section{Introduction}
\label{introgen}

Recent attempts to study the non--perturbative properties of
gauge and string theories have made an essential use of the low energy
supergravity lagrangians encoding their global and local symmetries.
In this analysis, the BPS saturated states play an important role, and
the interpretation of the classical solutions as states belonging to
the non--perturbative string  spectrum \cite{duffrep},\cite{kstellec}
has found strong support with the advent of D--branes \cite{Polchtasi},
allowing the direct construction of the BPS states.

In virtue of several non--perturbative dualities between different string
models, a given supergravity theory is truly specified by the dimension
of space--time, the number of unbroken supersymmetries and the massless
matter content.
\par
Each supergravity theory can be formulated in terms of a non--compact form of
the group $ G = E_{11-D}$ under which the p--forms in the theory transform in a
suitable representation. The scalar fields parametrize the coset $E_{11-D}/H$,
$H$ being the maximal subgroup of $E_{11-D}$.
It is now well known \cite{huto2}  that the discrete version $E_{11-D}(\ZZ)$,
is the U--duality group unifying S and T-dualities  and is supposed to be an
exact symmetry of non--perturbative  string theory.

From an abstract viewpoint, BPS saturated states are characterized by the
fact that they preserve a fraction of
the original supersymmetries. This  means  that there is a suitable
projection operator $\IP^2_{BPS} =\IP_{BPS}$ acting on the supersymmetry charge
$Q_{SUSY}$, such that:
\begin{equation}
\left(\IP_{BPS} \,Q_{SUSY} \right) \, \vert \, \mbox{BPS }>= 0\, .
\label{bstato}
\end{equation}
\par
Supersymmetry representation theory implies that it is actually an exact state
of non--perturbative string theory. Moreover,
the classical BPS state is by definition
an element of a {\it short supermultiplet} and, if supersymmetry is unbroken,
it cannot be renormalized to a {\it long supermultiplet}.
\par

Since the supersymmetry transformation rules  are  linear in the first
derivatives of the fields, eq.\eqn{bstato} is actually a {\it system of first
order differential equations} that must be combined with the second
order field equations of supergravity. The  solutions common to both
system of equations are the classical BPS saturated states.

This paper  investigates the most general BPS saturated black hole solutions of
$D=4$ supergravity \cite{malda},\cite{gensugrabh},\cite{feka}
preserving 1/2 or 1/4 of the  $N=8$ supersymmetry (thus,
in our case the scalar fields span the coset space $ E_{7(7)}/SU(8)$).
This is achieved along the same lines of  previous work on the 1/8 case
\cite{noi3},
where the content of dynamical
 fields and charges was identified, but an explicit solution was given
only by setting to zero the axion fields.
Here  we will find a complete solution
corresponding to a specific configuration of the scalar fields and
charges. This particular configuration is obtained by performing $G$ and
$H$ transformations in such a way that the quantized and central charges
are put in the normal frame. Once the solution is obtained, we can recover the
most general one by acting on the charges and the scalar fields by the
U-duality group.

In terms of the gravitino and dilatino physical fields $\psi_{A\mu}$,
$\chi_{ABC}$, $A,B,C=1,\ldots,8$, equation (\ref{bstato})
is equivalent to
\begin{equation}
\delta_\epsilon \psi_{A\mu}=\delta_\epsilon \chi_{ABC}=0
\label{kse}
\end{equation}
whose solution is given in terms of the Killing spinor $\epsilon_A(x)$
subject to the supersymmetry preserving condition
\begin{eqnarray}
\gamma^0 \,\epsilon_{a} & =&  \mbox{i}\, \IC_{ab}
\,  \epsilon^{b} \quad ; \quad a,b=1,\dots ,n_{max} \nonumber\\
\epsilon_{X} & =& 0; \quad X=n_{max},\dots ,8 \nonumber\\
A &=&\{a,X\}\nonumber
\label{ksc}
\end{eqnarray}
where $n_{max}$ is twice the number of unbroken supersymmetries.

In the present context,
eq.\eqn{kse} has two main features:
\begin{enumerate}
\item{It requires an efficient parametrization of the scalar field
sector.} The use of the rank $7$ Solvable Lie Algebra (SLA)
 $Solv_7$ associated with $E_{7(7)}/SU(8)$ is of great help in this problem.

\item{It breaks the original $SU(8)$ automorphism group of the
supersymmetry algebra to the subgroup
$\hat H=Usp( \,n_{max})\times SU(8-\,n_{max})\times U(1)$}
\end{enumerate}

\par
Indeed, SLA's  are quite useful in the construction of BPS saturated black hole
solutions since they provide us
with an efficient tool for decomposing the scalar sector
parametrizing $E_{7(7)}\over {SU(8)}$ in a way appropriate to the
decomposition of the isotropy subgroup $SU(8)$ with respect to the
subgroup $\hat H\subset SU(8)$ leaving invariant the Killing
condition on the supersymmetry parameters.

The relevant decomposition also yields an answer to the fundamental
question:
{\sl How many scalar fields are essentially dynamical, namely cannot
be set to constants up to U--duality transformations?}
\par

The answer is provided in terms of the SLA description of the scalar fields,
whose  evolution from arbitrary values at infinity to fixed values at the
horizon is best understood and dealt with when the scalars are algebraically
characterized as the parameters associated with the generators of the SLA.
Infact, an SLA  $Solv \left(  G/ H \right)$ is associated with every
non--compact homogeneous space and so in particular with $E_{7(7)}/SU(8)$. The
scalar fields are in one--to--one correspondence with the generators of the
SLA.

Once the charges are reduced to the normal frame, we consider the stabilizer
$G_{stab}\subset E_{7(7)}$ leaving the charge configuration invariant,
and its normalizer group $G_{norm}$.
Considering its maximal compact subgroup
$H_{norm} \subset G_{norm}$ we obtain a solvable Lie
algebra $Solv_{norm} \left( G_{norm}/H_{norm}\right)$ which is a
subalgebra of $Solv\left(E_{7(7)}/SU(8)\right)$. The corresponding
scalar fields are the essentially dynamical ones up to U--duality
transformations. The remaining scalar fields are either gauge--fixed to zero
by putting the vector of quantized charges into normal form or they belong to
$H_{norm}$ and are thus ineffective.

\par

After completing such decomposition in terms of SLA's we use the group
theoretical structure of the Killing spinor equations in order to decompose
them into fragments of the $SU(8)$ subgroup $\hat H$ . This analysis
is completely consistent with the results based on the SLA's and we are
left with a set of first order equations on the surviving scalar fields
and electromagnetic field strengths. At this point one uses the U-duality
group theoretical construction of the coset representative and the kinetic
matrix of the vectors explained in \cite{noi3} to compute the
geometrical objects
 appearing in the Killing spinor equations. This also allows us to compute
the truncated bosonic lagrangian for the model reduced to the essential
degrees of freedom and enjoying 1/2 or 1/4 of the original supersymmetry.
Coupling Killing spinor first order differential equation with the second
 order equations of motion we find solutions for both.
Actually the reduced lagrangian obtained in this way were already
known and well studied in the literature and our solution fit nicely in
the general scheme studied in ref \cite{popelu}.

The paper is organized as follows:

In section 2 we give a preliminary discussion of the $N=8$ BPS black
holes. Using the results of ref \cite{FGun} we discuss the stabilizer and normalizer
group for the charges reduced to normal form. Furthermore, we give a
the SLA decompositions in the 1/2 and 1/4 case and we quote the
results obtained from the detailed study of the Killing spinor
equations as far as the content of physical fields and charges is
concerned. For completeness we also discuss the analogous results
obtained in ref. \cite{noi3} in the 1/8 case.
\par
In section 3 we give a short resume` of the group theoretical
structure of $N=8$ supergravity in our conventions and the definition
of the central and quantized charges.
\par
In section 4 we analyse the decomposition of the Killing spinor
equations  under the group $\hat H$ in the 1/2 case and compute explicitly the
$U$- duality coset representative and the associated metric for the surviving vector field.
This enables us to recover the reduced Lagrangian and to identify it
as a well known model studied in the literature.Furthemore we also
give an alternative approach to the 1/2 case using the Dynkin
formalism for the determination of the relevant physical quantities.
\par
In section 5 the same kind of analysis is given for the 1/4 case:in
particular we show how our Lagrangian and solution fits in the so
called "$p$-brane taxonomy of ref. \cite{popelu}.
\par
In section 6 we give the conclusions.
\section{N=8 BPS black holes: a general discussion}
\label{generaldisc}

\par
The $D=4$ supersymmetry algebra with $N=8$  supersymmetry charges  is given by
\begin{eqnarray}
&\left\{ {\bar Q}_{A  \alpha }\, , \,{\bar Q}_{B  \beta}
\right\}\, = \,  {\rm i} \left( {\bf C} \, \gamma^\mu \right)_{\alpha \beta} \,
P_\mu \, \delta_{AB} \, \, - \, {\bf C}_{\alpha \beta} \,
\ZZ_{AB}& \nonumber\\
&\left( A,B = 1,\dots,8 \right)&
\label{susyeven}
\end{eqnarray}
where the SUSY charges ${\bar Q}_{A}\equiv Q_{A}^\dagger \gamma_0=
Q^T_{A} \, {\bf C}$ are Majorana spinors, $\bf C$ is the charge conjugation
matrix, $P_\mu$ is the 4--momentum operator and
 the antisymmetric tensor
$\ZZ_{AB}=-\ZZ_{BA}$ is the central charge operator. It
can always be reduced to normal form
\begin{equation}
\ZZ_{AB} ~~~~= \pmatrix{\epsilon Z_1&0&0&0\cr 0&\epsilon
Z_2&0&0\cr 0&0&\epsilon Z_3&0\cr 0&0&0&\epsilon Z_4\cr}
\end{equation}
where $\epsilon$ is the $2\times 2$ antisymmetric matrix,
(every zero is a $2\times2$ zero matrix) and
the four skew eigenvalues $Z_i$ of $\ZZ_{AB}$ are the central charges.
\par
Consider the reduced supercharges:
\begin{eqnarray}
{\bar S}^{\pm}_{A  \alpha }&=&\frac{1}{2} \,
\left( {\bar Q}_{A} \gamma _0 \pm \mbox{i} \,
\IC_{AB} \,  {\bar Q}_{B}\,
\right)_ \alpha ~~; ~~~~~~ A,B = 1,\dots, n_{max}\nonumber\\
 {\bar S}^{\pm}_{A  \alpha }&=& 0~~~~~ A > n_{max}
\label{redchar}
\end{eqnarray}

where  $\IC_{AB}$ is the invariant symplectic metric
($\IC=-\IC^T, \IC^2=-{\bf 1}$).
They can be regarded as the result of applying
a projection operator to the supersymmetry
charges: $ {\bar S}^{\pm}_{A}  =  {\bar Q}_{B} \, \IP^\pm_{BA} $,
where $ \IP^\pm_{BA} = \frac{1}{2}\, \left({\bf 1}\delta_{BA} \pm \mbox{i}
\IC_{BA} \gamma_0 \right)$. In the rest frame where the four momentum
is $P_\mu$ =$(M,0,0,0)$, we obtain the algebra:
$
\left\{ {\bar S}^{\pm}_{A}  \, , \, {\bar S}^{\pm}_{B} \right\} =
\pm \IC_{AC}\, {\bf C} \, \IP^\pm_{CB} \, \left( M \mp Z_I \right)\,
\delta_{IJ}
$ and the BPS states that saturate the bounds $
\left( M\pm Z_I \right) \,\vert \mbox{BPS,} i\rangle = 0
$ are those which are annihilated by the corresponding reduced supercharges:
\begin{equation}
{\bar S}^{\pm}_{A}   \, \vert \mbox{BPS ,} i\rangle = 0
\label{susinvbps}
\end{equation}
Eq.\eqn{susinvbps} defines {\sl short multiplet
representations} of the original algebra \eqn{susyeven} in the
following sense: one constructs a linear representation of \eqn{susyeven}
where all states are identically
annihilated by the operators ${\bar S}^{\pm}_{A}$ for $A=1,\dots,n_{max}$.
If $n_{max}=2$ we have the minimum shortening, if $n_{max}=8$ we
have the maximum shortening. On the other hand eq.\eqn{susinvbps}
can be translated into first order differential equations on the
bosonic fields of supergravity  whose common solutions with the ordinary
field  equations are the BPS saturated black hole configurations. In
the case of maximum shortening $n_{max}=8$ the black hole preserves
$1/2$ supersymmetry, in the case of intermediate shortening
$n_{max}=4$  it preserves $1/4$, while in the case of minimum
shortening it preserves  $1/8$.

\subsection{The Killing spinor equation and its covariance group }
In order to translate eq.\eqn{susinvbps} into first order differential equations
 on the bosonic fields of supergravity we consider
a configuration where all the fermionic fields are zero and we set
to zero the fermionic SUSY rules appropriate to such a background
\begin{equation}
0=\delta \mbox{fermions} =
\mbox{SUSY rule} \left( \mbox{bosons},\epsilon_{Ai} \right)
\label{fermboserule}
\end{equation}
and to a SUSY parameter that satisfies the following conditions:
\begin{equation}
\begin{array}{rclcl}
\xi^\mu \, \gamma_\mu \,\epsilon_{A} &=& \mbox{i}\, \IC_{AB}

\,  \epsilon^{B}   & ; &   A,B=1,\dots,n_{max}\\
\epsilon_{A} &=& 0  &;&   A> n_{max} \\
\end{array}
\label{kilspieq}
\end{equation}
Here $\xi^\mu$ is a time--like Killing vector for the space--time metric
( in the following we just write
 $\xi^\mu \gamma_{\mu} = \gamma^0$) and
$ \epsilon _{A}, \epsilon^{A}$ denote the two chiral projections of
a single Majorana spinor: $ \gamma _5 \, \epsilon _{A} \, = \, \epsilon _{A} $ ,
$ \gamma _5 \, \epsilon ^{A} \, = - \epsilon ^{A} $
We name eq.\eqn{fermboserule} the {\sl Killing spinor equation}
and the investigation of its
group--theoretical structure was our main goal in ref \cite{noi3}.
There we restricted our attention to the case $n_{max}=2$: here
we considere the other two possibilities.


To appreciate the distinction among the three types of $N=8$ black--hole solutions
we need to recall the results of \cite{FGun} where a classification was given of the
${\bf 56}$--vectors of quantized electric and magnetic charges ${\vec Q}$ characterzing
such solutions. The basic argument is provided by the reduction of
the central charge skew--symmetric tensor $\ZZ_{AB}$ to normal form.
The reduction can always be obtained by means of local $SU(8)$
transformations, but the structure of the skew eigenvalues depends on
the orbit--type of the $\bf 56$--dimensional charge vector which can be described by means
of its stabilizer subgroup $G_{stab}({\vec Q}) \subset E_{7(7)}$:
\begin{equation}
g \, \in \, G_{stab}({\vec Q}) \subset E_{7(7)} \quad \Longleftrightarrow \quad
g \,  {\vec Q}  = {\vec Q}
\end{equation}
There are three possibilities:
\begin{equation}
\begin{array}{cccc}
\null & \null & \null & \null \\
 \mbox{SUSY} & \mbox{Central Charge} & \mbox{Stabilizer}\equiv G_{stab} &
 \mbox{Normalizer}\equiv G_{norm}\\
  \null & \null & \null & \null \\
   1/2 & Z_1 = Z_2 =Z_3 =Z_4  & E_{6(6)} & O(1,1) \\
 \null & \null & \null & \null \\
 1/4 & Z_1 = Z_2 \neq Z_3=Z_4 & SO(5,5)&
 SL(2,\IR) \times O(1,1) \\
 \null & \null & \null & \null \\
 1/8 & Z_1 \ne Z_2 \ne Z_3 \ne Z_4 & SO(4,4) & SL(2,\IR)^3 \\
 \null & \null & \null & \null \\
\label{classif}
\end{array}
\end{equation}
where the normalizer $G_{norm}({\vec Q})$ is defined as the subgroup
of $E_{7(7)}$ that commutes with the stabilizer:
\begin{equation}
   \left[ G_{norm} \, , \,  G_{stab} \right] = 0
\end{equation}
The main result of \cite{noi3} is that the most general $1/8$
black--hole solution of $N=8$ supergravity is related to the normalizer group
$SL(2,\IR)^3$. In this paper we show that analogous relations
apply also to the other cases.

\subsubsection{The $1/2$ SUSY case}
Here we have $n_{max} = 8$ and correspondingly the covariance subgroup
of the Killing spinor equation is $Usp(8)\, \subset \, SU(8)$. Indeed
 condition \eqn{kilspieq} can be rewritten as follows:
 \begin{equation}
\label{1/2killingspin}
\gamma^0 \,\epsilon_{A}  =  \mbox{i}\, \IC_{AB}
\,  \epsilon^{B} \quad ; \quad A,B=1,\dots ,8
\label{urcunmez}
\end{equation}
where $\IC_{AB}= - \IC_{BA}$ denotes an $ 8 \times 8$
antisymmetric matrix satisfying $\IC^2 = -\bfone$. The group
$Usp(8)$ is the subgroup of unimodular, unitary $ 8 \times 8$
matrices that are also symplectic, namely that preserve the matrix
$\IC$. Relying on eq. (\ref{classif}) we see that in the present case $G_
{\it stab}=E_{6(6)}$ and $G_{\it norm}=O(1,1)$. Furthermore we have the
following
decomposition of the ${\bf
70}$ irreducible representation of $SU(8)$ into irreducible
representations of $Usp(8)$:
\begin{equation}
{\bf 70} \, \stackrel{Usp(8)}{\longrightarrow} \, {\bf 42} \, \oplus
\, {\bf 1} \, \oplus \, {\bf 27}
\label{uspdecompo1}
\end{equation}
\par
We are accordingly lead to decompose the solvable Lie algebra as
\begin{eqnarray}
Solv_7 & =& Solv_6 \, \oplus \, O(1,1) \, \oplus \ID_6
\label{decomp1a}\\
70 & = & 42 \, + \, 1 \, + \, 27
\label{decomp1b}
\end{eqnarray}
where, following the notation established in \cite{noi1,noi2}:
\begin{eqnarray}
Solv_7 & \equiv & Solv \left(\frac{E_{7(7)}}{SU(8)}\right)
\nonumber\\
Solv_6 & \equiv & Solv \left(\frac{E_{6(6)}}{Usp(8)}\right)
\nonumber\\
\mbox{dim}\, Solv_7 & = & 70 \quad ; \quad \mbox{rank}\, Solv_7 \, =
\, 7 \nonumber\\
\mbox{dim}\, Solv_6 & = & 42 \quad ; \quad \mbox{rank}\, Solv_6 \, =
\, 6 \nonumber\\
\label{defsolv7}
\end{eqnarray}
In eq.\eqn{decomp1a}
$Solv_6$ is the solvable Lie algebra that describes the scalar
sector of $D=5$, $N=8$ supergravity, while the $27$--dimensional
abelian ideal $\ID_6$ corresponds to those $D=4$ scalars that
originate from the $27$--vectors of supergravity one--dimension
above \cite{noi2}.
Furthermore, we may also decompose the $\bf 56$ charge representation
of $E_{7(7)}$ with respect to $O(1,1)\times E_{6(6)}$ obtaining
\begin{equation}
{\bf 56}\stackrel{Usp(8)}{\longrightarrow}
({\bf 1,27})\oplus({\bf 1,27})\oplus({\bf 2,1})
\label{visto}
\end{equation}

In order to  single out the content of the first order Killing
spinor equations we need to decompose them into irreducible $Usp(8)$
representations.
The gravitino equation is an ${\bf 8}$ of $SU(8)$ that
remains irreducible under $Usp(8)$ reduction. On the other hand the dilatino
equation is a ${\bf 56}$ of $SU(8)$ that reduces as follows:
\begin{equation}
{\bf 56} \, \stackrel{Usp(8)}{\longrightarrow} \, {\bf 48} \, \oplus
\, {\bf 8}
\label{uspdecompo2}
\end{equation}
Hence altogether we have that $3$ Killing spinor equations in the
representations ${\bf 8}$,
${\bf 8}^\prime$ , ${\bf 48}$ constraining the scalar fields
 parametrizing  the three subalgebras
${\bf 42}$, ${\bf 1}$ and ${\bf 27}$. Working out
the consequences of these constraints and deciding which scalars are
set to constants, which are instead evolving and how many charges
are different from zero is the content of
section \ref{det1/2}.    As it will be seen explicitly there
the
content of the Killing spinor equations after $Usp(8)$ decomposition,
is such as to set to a constant $69$ scalar fields parametrizing
$Solv_6\oplus\ID_6$ thus confirming the SLA analysis discussed in the
introduction: indeed in this case $G_{norm} = O(1,1)$ and $H_{norm}
=\bf 1$, so that there is just one
 surviving field parametrising $G_{norm}= O(1,1)$.
Moreover, the same Killing spinor equations
tell us that the $54$ belonging to the two $({\bf 1},{\bf 27})$
representation of eq. (\ref{visto})
are actually zero, leaving only two non--vanishing charges transforming
as a doublet of $O(1,1)$.

\subsubsection{The $1/4$ SUSY case}
Here we have $n_{max} = 4$ and correspondingly the covariance subgroup
of the Killing spinor equation is $Usp(4)\,\times \, SU(4) \,\times \, U(1)
\subset \, SU(8)$. Indeed condition \eqn{kilspieq} can be rewritten as follows:
\begin{eqnarray}
\gamma^0 \,\epsilon_{a} & =&  \mbox{i}\, \IC_{ab}
\,  \epsilon^{b} \quad ; \quad a,b=1,\dots ,4 \nonumber\\
\epsilon_{X} & =& 0; \quad X=5,\dots ,8 \nonumber\\
\label{urcunquart}
\end{eqnarray}
where $\IC_{ab}= - \IC_{ba}$ denotes a  $ 4 \times 4$
antisymmetric matrix satisfying $\IC^2 = -\bfone$. The group
$Usp(4)$ is the subgroup of unimodular, unitary $ 4 \times 4$
matrices that are also symplectic, namely that preserve the matrix
$\IC$.
\par
We are accordingly lead to decompose the solvable Lie algebra as follows.
Naming:
\begin{eqnarray}
Solv_S & \equiv & Solv \left(\frac{SL(2,R)}{U(1)}\right)
\nonumber\\
Solv_T & \equiv & Solv \left(\frac{SO(6,6)}{SO(6) \times SO(6)}\right)
\nonumber\\
\mbox{dim}\, Solv_S & = & 2 \quad ; \quad \mbox{rank}\, Solv_S \, =
\, 1 \nonumber\\
\mbox{dim}\, Solv_T & = & 36 \quad ; \quad \mbox{rank}\, Solv_T \, =
\, 6 \nonumber\\
\label{defsolst}
\end{eqnarray}
we can write:
\begin{eqnarray}
Solv_7 & =& Solv_S \, \oplus \, Solv_T \, \oplus \, {\cal W}_{32}
\label{stdecomp1a}\\
70 & = & 2\, + \, 36 \, + \, 32  \ .
\label{stdecomp1b}
\end{eqnarray}
 As shown in ref. \cite{noi1},\cite{noi2},
 the SLA's $Solv_S$ and $Solv_T$ describe the dilaton--axion
sector and the six torus moduli, respectively, in the interpretation
of $N=8$ supergravity as the compactification of Type IIA theory on a six--torus
$T^6$ \cite{noi2}. The rank zero abelian subalgebra ${\cal W}_{32}$
is instead composed by  Ramond-Ramond scalars.
\par
Introducing the decomposition \eqn{stdecomp1a}, \eqn{stdecomp1b} we have
succeeded in singling out a holonomy subgroup $SU(4) \, \times \,
SU(4) \, \times \, U(1) \, \subset \, SU(8)$. Indeed we have
$SO(6) \, \equiv \, SU(4)$. This is a step forward but it is not yet
the end of the story since we actually need a subgroup $Usp(4) \,
\times \, SU(4)\times U(1)$ corresponding to the invariance group of equation
(\ref{1/2killingspin}). This means that we must further decompose the
SLA $Solv_T$. This latter is the manifold of the
scalar fields associated with vector multiplets in an $N=4$
decomposition of the $N=8$ theory. Indeed the decomposition
\eqn{stdecomp1a} with respect to the S--T--duality subalgebra is the
appropriate decomposition of the scalar sector according to $N=4$
multiplets.

The  further SLA decomposition we need is  :
\begin{eqnarray}
Solv_T &=& Solv_{T5} \oplus Solv_{T1}  \nonumber\\
Solv_{T5}& \equiv &Solv \left( \frac{SO(5,6)}{SO(5) \times
SO(6)}\right)\nonumber  \\
Solv_{T1}& \equiv &Solv \left( \frac{SO(1,6)}{
SO(6)}\right)\nonumber  \\
\label{stspilt}
\end{eqnarray}
where we rely on the
isomorphism $Usp(4)\equiv SO(5)$ and we have taken into
account that the $\bf{70}$ irreducible
representation of $SU(8)$ decomposes with respect to
$Usp(4) \, \times \, SU(4)\times U(1)$ as follows

\begin{equation}
{\bf 70} \, \stackrel{Usp(4) \, \times \, SU(4) \, \times \, U(1)}{\longrightarrow}
\, \left( {\bf 1},{\bf 1},{\bf 1}+{\bar {\bf 1}}\right) \, \oplus
\, \left({\bf 5},{\bf 6}, {\bf 1} \right) \, \oplus \,
\left( {\bf 1},{\bf 6},{\bf 1} \right) \, \oplus \,
\left( {\bf 4} ,{\bf 4}, {\bf 1}\right) \, \oplus \,
\left( {\bf 4},{\bf 4},{\bf 1} \right)
\label{uspdecompo3}
\end{equation}
Hence,
altogether we can write:
\begin{eqnarray}
Solv_7 &=& Solv_S \, \oplus \, Solv_{T5} \, \oplus \, Solv_{T1} \,
\oplus \, {\cal W}_{32} \nonumber\\
70 & = & 2 \, + \, 30 \, + \,  6\, + 32
\end{eqnarray}

Just as in the previous case we should now
single out the content of the first order Killing
spinor equations by decomposing  them into irreducible
$Usp(4) \, \times \, SU(4) \, \times \, U(1)$ representations.
The dilatino equations $\delta \chi_{ABC}=0$, and the gravitino
equation $\delta\psi_A=0$, $A,B,C=1,\ldots,8$  ($SU(8)$ indices),
decompose as follows
\begin{equation}
{\bf 56} \, \stackrel{Usp(4) \, \times \, SU(4) }{\longrightarrow}
\, ( {\bf 4},{\bf 1}) \, \oplus
\, 2({\bf 1},{\bf 4},)\, \oplus \,
( {\bf 5},{\bf 4}) \oplus \,
( {\bf 4} ,{\bf 6})
\label{uspdecomp56}
\end{equation}
\begin{equation}
{\bf 8} \, \stackrel{Usp(4) \, \times \, SU(4) }{\longrightarrow}
\, ( {\bf 4},{\bf 1}) \, \oplus
\, ({\bf 1},{\bf 4},)
\label{usp56}
\end{equation}

As we
shall see explicitly in
section \ref{seckilling1/4}, the content of the reduced Killing
spinor equations is such that only two scalar fields are essentially
dynamical all the other being set to constant up to $U$- duality
transformations. Moreover $52$ charges are set to zero leaving $4$
charges transforming the $(2,2)$ representation of $Sl(2,\IR)\times
O(1,1)$.
Note that in the present case on the basis of the SLA analysis given
in the introduction, one would expect $3$ scalar fields parametrising
 $ G_{norm}/ H_{norm}$ =
${Sl(2,\IR)\over U(1)} \times O(1,1)$; however the relevant Killing
spinor equation gives an extra reality constraint on the
$Sl(2,\IR)\over U(1)$ field
thus reducing the number of non trivial scalar fields to two.
\subsubsection{The $1/8$ SUSY case}
Here we have $n_{max} = 2$ and
$Solv_7$ must be decomposed  according to the decomposition
of the isotropy subgroup: $SU(8) \longrightarrow SU(2)\times U(6)$. We
showed in \cite{noi3}  that the corresponding decomposition of the solvable
Lie algebra is the following one:
\begin{equation}
Solv_7   =  Solv_3 \, \oplus \, Solv_4
\label{7in3p4}
\end{equation}
\begin{equation}
\begin{array}{rclrcl}
Solv_3 & \equiv & Solv \left( SO^\star(12)/U(6) \right) & Solv_4 &
\equiv & Solv \left( E_{6(4)}/SU(2)\times SU(6) \right) \\
\mbox{rank }\, Solv_3 & = & 3 & \mbox{rank }\,Solv_4 &
= & 4 \\
\mbox{dim }\, Solv_3 & = & 30 & \mbox{dim }\,Solv_4 &
= & 40 \\
\end{array}
\label{3and4defi}
\end{equation}
The rank three  Lie algebra $Solv_3$ defined above describes the
thirty dimensional scalar sector of $N=6$ supergravity, while the rank four
solvable Lie algebra $Solv_4$ contains the remaining forty scalars
belonging to $N=6$ spin $3/2$ multiplets. It should be noted
that, individually, both manifolds $ \exp \left[ Solv_3 \right]$ and
$ \exp \left[ Solv_4 \right]$ have also an $N=2$ interpretation since we have:
\begin{eqnarray}
\exp \left[ Solv_3 \right] & =& \mbox{homogeneous special K\"ahler}
\nonumber \\
\exp \left[ Solv_4 \right] & =& \mbox{homogeneous quaternionic}
\label{pincpal}
\end{eqnarray}
so that the first manifold can describe the interaction of
$15$ vector multiplets, while the second can describe the interaction
of $10$ hypermultiplets. Indeed if we decompose the $N=8$ graviton
multiplet in $N=2$ representations we find:
\begin{equation}
\mbox{N=8} \, \mbox{\bf spin 2}  \,\stackrel{N=2}{\longrightarrow}\,
 \mbox{\bf spin 2} + 6 \times \mbox{\bf spin 3/2} + 15 \times \mbox
{\bf vect. mult.}
 +
 10 \times \mbox{\bf hypermult.}
 \label{n8n2decompo}
\end{equation}
Introducing the decomposition \eqn{7in3p4}
we found in \cite{noi3} that the $40$ scalars belonging
to $Solv_4$ are constants
independent of the radial variable $r$. Only the $30$ scalars in the
K\"ahler algebra $Solv_3$ can be radial dependent. In fact their
radial dependence is governed by a first order differential equation
that can be extracted from a suitable component of the Killing spinor
equation.The result in this case is that $64$ of the scalar fields are
 actually constant
while $6$ are dynamical Moreover $48$ charges are annihilated leaving
$6$ nonzero chargs transforming in the representation $(2,2,2)$ of
the normalizer $G_{norm} = [Sl(2,\IR)]^3$.
More precisely we obtained the following result.
Up to U--duality transformations the most general $N=8$ black--hole
is actually an $N=2$ black--hole corresponding to a very
specific choice of the special K\"ahler manifold, namely $ \exp[ Solv_3 ]$
as in eq.\eqn{pincpal},\eqn{3and4defi}. Furthermore up to the duality
rotations of $SO^\star(12)$ this general solution is actually
determined by the so called $STU$ model studied in \cite{STUkallosh}
and based on the solvable subalgebra:
\begin{equation}
Solv \left( \frac{SL(2,\IR)^3}{U(1)^3} \right) \, \subset \, Solv_3
\label{rilevanti}
\end{equation}
\par
In other words the only truly indipendent degrees of freedom of the
black hole solution are given by three complex scalar fields,
$S,T,U$.
The real parts of these scalar fields correspond to the three Cartan
generators of $Solv_3$ and have the physical interpretation of radii
of the torus compactification from $D=10$ to $D=4$. The imaginary
parts of these complex fields are generalised theta angles.
\section{Summary of $N=8$ supergravity.}
\label{41}
We first establish the relevant definition and notation (for further
details see \cite{amicimiei}, \cite{noi3} and for a general review on the duality
formalism \cite{mylecture})
We introduce the coset representative $\IL$ of
$E_{\left(7\right)7}\over SU\left(8\right)$ in the ${\bf 56}$ representation
of $E_{\left(7\right)7}$:
\begin{equation}
\label{Lcoset}
\IL={1 \over\sqrt{2}}\left(\begin{array}{c|c}
                             f+\mbox{\rm i} h & \bar{f}+\mbox{\rm i}\bar{h}\\
                             \hline\\
                             f-\mbox{\rm i} h & \bar{f}-\mbox{\rm i}\bar{h}
                             \end{array}
                     \right)
\end{equation}
where the submatrices $\left(h,f\right)$ are $28\times28$
matrices indexed by antisymmetric pairs $\Lambda,\Sigma, A,B$
$\left(\Lambda,\Sigma=1,\dots,8;~~~A,B=1,\dots,8\right)$ the first
pair transforming under $E_{\left(7\right)7}$ and the second one under
$SU\left(8\right)$:
\begin{equation}
\left(h,f\right)=\left(h_{\Lambda\Sigma|AB},
f^{\Lambda\Sigma}_{~~AB}\right)
\end{equation}
Note that $\IL\in Usp\left(28,28\right)$.
The vielbein $P_{ABCD}$ and the $SU\left(8\right)$ connection
$\Omega_{A}^{~B}$ of $E_{\left(7\right)7}\over SU\left(8\right)$ are
computed from the left invariant $1$-form $\IL^{-1}d\IL$:
\begin{equation}
\label{L-1dL}
\IL^{-1}d\IL=\left(\begin{array}{c|c}
                             \delta^{[A}_{~~[C}\Omega^{B]}_{~~D]} & \bar{P}^{ABCD}\\
                             \hline\\
                             P_{ABCD} & \delta_{[A}^{~~[C}\bar{\Omega}_{B]}^{~~D]}\end{array}
                     \right)
\end{equation}
where $P_{ABCD}\equiv
P_{ABCD,\alpha}d\Phi^{\alpha}~~~\left(\alpha=1,\dots,70\right)$ is
completely antisymmetric and satisfies the reality condition
\begin{equation}
\label{realcondP}
P_{ABCD}={1 \over 24}\epsilon_{ABCDEFGH}\bar P^{EFGH}
\end{equation}
The bosonic lagrangian of $N=8$ supergravity is \cite{cre}
\begin{eqnarray}
\label{lN=8}
{\cal L}&=&\int\sqrt{-g}\, d^4x\left(2R+\Im{\cal
N}_{\Lambda\Sigma|\Gamma\Delta}F_{\mu\nu}^{~~\Lambda\Sigma}F^{\Gamma\Delta|\mu\nu}+
{1 \over 6}P_{ABCD,i}\bar{P}^{ABCD}_{j}\partial_{\mu}
\Phi^{i}\partial^{\mu}\Phi^{j}+\right.\nonumber\\
&+&\left.{1 \over 2}\Re{\cal N}_{\Lambda\Sigma|\Gamma\Delta}
{\epsilon^{\mu\nu\rho\sigma}\over\sqrt{-g}}
F_{\mu\nu}^{~~\Lambda\Sigma}F^{\Gamma\Delta}_{~~\rho\sigma}\right)
\end{eqnarray}
where the curvature two-form is defined as
\begin{equation}
R^{ab}=d\omega^{ab}-\omega^a_{~c}\wedge\omega^{cb}.
\end{equation}
and the kinetic matrix ${\cal N}_{\Lambda\Sigma|\Gamma\Delta}$ is
given by:
\begin{equation}
\label{defN}
{\cal N}=hf^{-1}~~\rightarrow~~ {\cal N}_{\Lambda\Sigma|\Gamma\Delta}=
h_{\Lambda\Sigma|AB}f^{-1~AB}_{~~~~~\Gamma\Delta}.
\end{equation}
The same matrix relates the (anti)self-dual electric and magnetic
$2$-form field strengths, namely, setting
\begin{equation}
\label{defFpm}
F^{\pm~\Lambda\Sigma}={1 \over 2}\left(F\pm \mbox{\rm i}~\star
F\right)^{\Lambda\Sigma}
\end{equation}
one has
\begin{eqnarray}
G^{-}_{\Lambda\Sigma}&=&\bar{{\cal N}}_{\Lambda\Sigma|\Gamma\Delta}F^{-~\Gamma\Delta}
\nonumber\\
\label{defG}
G^{+}_{\Lambda\Sigma}&=&{\cal N}_{\Lambda\Sigma|\Gamma\Delta}F^{+~\Gamma\Delta}
\end{eqnarray}
where the ''dual'' field strengths $G^{\pm}_{\Lambda\Sigma}$ are
defined as $G^{\pm}_{\Lambda\Sigma}={i \over 2}{\delta{\cal L} \over
\delta F^{\pm~\Lambda\Sigma}}$.
Note that the $56$ dimensional (anti)self-dual vector $\left(F^{\pm~\Lambda\Sigma},G^{\pm}_{\Lambda\Sigma}\right)$
transforms covariantly under $U\in Sp\left(56,\IR\right)$
\begin{eqnarray}
&&U\left(\begin{array}{c} F \\ G\end{array}\right)=\left(\begin{array}{c} F' \\ G'\end{array}\right)~~;~~
U=\pmatrix{A&B\cr C&D\cr}\nonumber\\
&&A^tC-C^tA=0\nonumber\\
&&B^tD-D^tB=0\nonumber\\
&&A^tD-C^tB={\bf 1}
\end{eqnarray}
The matrix transforming the coset representative $\IL$ from the
$Usp\left(28,28\right)$ basis, eq.\eqn{Lcoset}, to the real
$Sp\left(56,\IR\right)$ basis is the Cayley matrix:
\begin{equation}
\IL_{Usp}={\cal C}\IL_{Sp}{\cal C}^{-1}~~~~~{\cal C}=\pmatrix{ \bfone &
\mbox{\rm i} \bfone
\cr \bfone &-\mbox{\rm i}\bfone \cr}
\label{caylone}
\end{equation}
implying
\begin{eqnarray}
f&=&{1 \over\sqrt{2}}\left(A-\mbox{\rm i} B\right)\nonumber\\
h&=&{1 \over\sqrt{2}}\left(C-\mbox{\rm i} D\right)
\end{eqnarray}
Having established our definitions and notations, let us now write
down the Killing spinor equations obtained by equating to zero the
SUSY transformation laws of the gravitino $\psi_{A\mu}$ and dilatino $\chi_{ABC}$ fields
of $N=8$ supergravity:
\begin{eqnarray}
\label{chisusy}
\delta\chi_{ABC}&=&~4 \mbox{\rm i} ~P_{ABCD|i}\partial_{\mu}\Phi^i\gamma^{\mu}
\epsilon^D-3T^{(-)}_{[AB|\rho\sigma}\gamma^{\rho\sigma}
\epsilon_{C]}+\dots=0\\
\label{psisusy}
\delta\psi_{A\mu}&=&\nabla_{\mu}\epsilon_A~
~-\frac{\rm 1}{4}\, T^{(-)}_{AB|\rho\sigma}\gamma^{\rho\sigma}\gamma_{\mu}\epsilon^{B}+\dots=0
\end{eqnarray}
 where the dots denote unessential trilinear fermion terms,
$\nabla_{\mu}$ denotes the derivative covariant both with respect to
Lorentz and $SU\left(8\right)$ local transformations
\begin{equation}
\nabla_{\mu}\epsilon_A=\partial_{\mu}\epsilon_A-{1\over
4}\gamma_{ab} \, \omega^{ab}\epsilon_A-\Omega_A^{~B}\epsilon_B
\label{delepsi}
\end{equation}
and where $T^{\left(-\right)}_{AB}$ is the ''dressed graviphoton''$2-$form,
  defined
as follows
\begin{equation}
T^{\left(-\right)}_{AB}=\left(h_{\Lambda\Sigma AB}
\left(\Phi\right)F^{-\Lambda\Sigma}-f^{\Lambda\Sigma}_{~~AB}
\left(\Phi\right)G^-_{\Lambda\Sigma}\right)
\label{tab}
\end{equation}
From equations (\ref{defN}), (\ref{defG}) we have the following
identities:
\begin{eqnarray}
&&T^+_{AB}=0\rightarrow T^-_{AB}=T_{AB}\nonumber ~~~~~~
\bar{T}^-_{AB}=0\rightarrow \bar{T}^+_{AB}=\bar{T}_{AB}
\end{eqnarray}
Note that the central charge is defined as
\begin{equation}
Z_{AB}=\int_{S^2}{T_{AB}}=h_{\Lambda\Sigma|AB}g^{\Lambda\Sigma}-f^{\Lambda\Sigma}_{~~AB}e_{\Lambda\Sigma}
\label{zab}
\end{equation}
where the integral of the two-form $T_{AB}$ is evaluated on a large
two-sphere at infinity and the quantized charges ($e_{\Lambda\Sigma},~g^{\Lambda\Sigma}$) are defined by
\begin{eqnarray}
g^{\Lambda\Sigma}&=&\int_{S^2}{F^{\Lambda\Sigma}}\nonumber\\
e_{\Lambda\Sigma}&=&\int_{S^2}{\cal N}_{\Lambda\Sigma|\Gamma\Delta}\star F^{\Gamma\Delta}.
\end{eqnarray}

\section{Detailed study of the $1/2$ case}

\label{det1/2}
As established in  \cite{FGun}, the $N=1/2$ SUSY preserving black hole
solution of $N=8$ supergravity has $4$ equal skew eingenvalues
in the normal frame for the central charges.
The stabilizer of the normal form is $E_{\left(6,6\right)}$
and the normalizer of this latter in $E_{7(7)}$ is $O\left(1,1\right)$:
\begin{equation}
E_{\left(7,7\right)}\supset E_{\left(6,6\right)}\times O\left(1,1\right)
\label{orfan1}
\end{equation}
According to our previous discussion, the relevant subgroup
of the $SU\left(8\right)$ holonomy group is $Usp\left(8\right)$, since
the BPS Killing spinor conditions involve supersymmetry parameters
$\epsilon_A,~\epsilon^A$ satisfying eq.\eqn{urcunmez}.
Relying on this information, we can write the solvable Lie algebra
decomposition \eqn{decomp1a},\eqn{decomp1b}
of the $\sigma-$model scalar coset $E_{\left(7,7\right)}
\over SU\left(8\right)$.
\par

As discussed in the introduction, it is natural to guess that modulo $U-$duality
transformations the complete solution is given in terms of a single scalar field
parametrizing $O\left(1,1\right)$.

Indeed, we can  now demonstrate that according to the previous
discussion there is just one
 scalar field, parametrizing the normalizer $O\left(1,1\right)$, which appear in the
final lagrangian, since the Killing spinor equations imply that
$69$ out of the $70$ scalar fields are actually  constants.
In order to achieve this result, we have to decompose the $SU\left(8\right)$ tensors
appearing in the equations (\ref{chisusy}),(\ref{psisusy}) with respect to $Usp\left(
8\right)$ irreducible representations. According to the decompositions
\begin{eqnarray}
\label{70-28decomp1/2}
{\bf 70}& \stackrel{Usp(8)}{=} &{\bf 42}\oplus{\bf 27}\oplus{\bf 1}\nonumber\\
{\bf 28}& \stackrel{Usp(8)}{=} &{\bf 27}\oplus{\bf 1}
\end{eqnarray}
we have
\begin{eqnarray}
\label{usp8dec}
P_{ABCD}&=&\stackrel{\circ}{P}_{ABCD}+{3\over 2}
C_{[AB}\stackrel{\circ}{P}_{CD]}+{1\over 16}C_{[AB}C_{CD]}P\nonumber\\
T_{AB}&=&\stackrel{\circ}{T}_{AB}+{1\over 8}C_{AB}T
\end{eqnarray}
where the notation $ \stackrel{\circ}{t}_{A_1 \dots , A_n}$
means that the antisymmetric tensor is  $Usp\left(8\right)$ irreducible, namely
has vanishing $C$-traces: $C^{A_1 A_2} \,\stackrel{\circ}{t}_{A_1 A_2 \dots , A_n}=0$.

Starting from  equation (\ref{chisusy}) and using equation (\ref{1/2killingspin})
we easily find:
\begin{equation}
4  P_{,a}\gamma^a\gamma^0-6T_{ab} \gamma^{ab}=0\, ,
\end{equation}
where we have twice contracted the free $Usp(8)$ indices with the
$Usp(8)$ metric $C_{AB}$. Next, using the decomposition
(\ref{usp8dec}), eq. (\ref{chisusy}) reduces to
\begin{equation}
-4  \left(  \stackrel{\circ}{P}_{ABCD,a}+ {3\over2} \stackrel{\circ}{P}_{[CD,a}C_{AB]}\right)
C^{DL}\gamma^a\gamma^0-3
\stackrel{\circ}{T}_{[AB}\delta^L_{C]}\gamma^{ab}=0\, .
\label{basta}
\end{equation}
Now we may alternatively contract equation (\ref{basta}) with
$C^{AB}$ or $\delta^L_C$ obtaining two relations on $ \stackrel{\circ}{P}_{AB} $
and   $\stackrel{\circ}{T}_{AB}$ which imply that they are separately
zero:
\begin{equation}
\stackrel{\circ}{P}_{AB}=\stackrel{\circ}{T}_{AB}=0\, ,
\end{equation}
which also imply, taking into account (\ref{basta})
\begin{equation}
\stackrel{\circ}{P}_{ABCD}=0\, .
\end{equation}
Thus
we have  reached the conclusion
\begin{eqnarray}
\label{42-27=0}
\stackrel{\circ}{P}_{ABCD|i}\partial_{\mu}\Phi^i\gamma^{\mu}\epsilon^D&=&0\nonumber\\
\stackrel{\circ}{P}_{AB|i}\partial_{\mu}\Phi^i\gamma^{\mu}\epsilon^B&=&0\\
\stackrel{\circ}{T}_{AB}&=&0 \label{tnot}
\end{eqnarray}
implying that $69$ out the $70$ scalar fields are actually
constant, while the only surviving central charge is that
associated with the singlet two-form $T$.Since $T_{AB}$ is a complex
combination of the electric and magnetic field strengths \ref{tab},
it is clear that eq. \ref{tnot} implies the vanishing of $54$ of the
quantized charges $g^{\Lambda \Sigma},e_{\Lambda \Sigma}$,the
surviving two charges trasforming as a doublet of $O(1,1)$ according
to eq. (\ref{visto}).
The only non-trivial evolution equation relates $P$ and $T$ as follows:
\begin{equation}
\label{singleteqn1/2}
\left(
\hat P\partial_{\mu}\Phi\gamma^{\mu}-{3\over 2}
\mbox{\rm i}\, T^{(-)}_{\rho\sigma}\gamma^{\rho\sigma}\gamma^0 \right)\epsilon_A=0
\end{equation}
where we have set $P = \hat P d\Phi $  and $\Phi$ is the unique non trivial scalar
field parametrizing $O(1,1)$.
\par
In order to make this equation explicit we perform the usual static ansaetze: \\
{\underline {\sl Black hole metric:}}
\begin{equation}
\label{ansaetze}
ds^2=e^{2{\cal U}\left(r\right)}dt^2-e^{-2{\cal
U}\left(r\right)}d\vec{x}^2 ~~~~~\left(r^2=\vec{x}^2\right)
\end{equation}
{\underline {\sl Matter fields:}}
\begin{eqnarray}
\Phi&=&\Phi\left(r\right)\\
F^{-\Lambda\Sigma}&=&{1 \over 4\pi}t^{\Lambda\Sigma}\left(r\right)E^{\left(-\right)}\label{F}\\
E^{\left(-\right)}&=&i {e^{2{\cal U}}\over r^3}dt\wedge x^idx^i+{x^i\over 2r^3}dx^j
\wedge dx^k\epsilon_{ijk}\label{e}\\
\label{smallt}
t^{\Lambda\Sigma}\left(r\right)&=&2\pi\left(g+\mbox{\rm i}q\left(r\right)\right)^{\Lambda\Sigma}
\end{eqnarray}
Using the definitions (\ref{defFpm}), (\ref{defG}), (\ref{tab}),(\ref{e}),(\ref{F}) we have
\begin{equation}
\label{Tab}
T_{ab}^- = {\rm i}\ t^{\Lambda\Sigma}(r) E_{ab}^- C^{AB}
{\rm Im}{\cal N }_{\Lambda \Sigma,\Gamma\Delta}f^{\Gamma\Delta}_{~~AB} \, .
\end{equation}
A simple gamma matrix manipulation gives further
\begin{equation}
\label{gammaalgebra}
\gamma_{ab}\, E^{\mp}_{ab}=
2 \mbox{\rm i} {e^{2{\cal U}}\over r^3}x^i\gamma^0\gamma^i\left(\pm 1+\gamma_5\over 2\right)
\end{equation}
and we arrive at the final equation
\begin{equation}
\label{dphidr}
\frac{d \Phi}{dr}= -\frac{\sqrt{3}}{4 }
q(r)^{\Lambda\Sigma}\, {\rm Im}{\cal N }_{\Lambda\Sigma|\Gamma\Delta}\,
f^{\Gamma\Delta}_{~~AB}\, \frac{e^{\cal U}}{r^2}\, .
\label{finale}
\end{equation}
In eq. (\ref{finale}), we have set $g^{\Lambda\Sigma}=0$ since
reality of  the l.h.s. and of
$f_{AB}^{\Gamma\Delta} $(see eq. (\ref{f})) imply the vanishing of the
magnetic charge.
Furthermore,
we have normalized the vielbein component of the $Usp(8)$ singlet
 as follows
\begin{equation}
\label{P}
\hat P=4 \sqrt{3}
\end{equation}
which corresponds to normalizing the $Usp(8)$ vielbein as
\begin{equation}
\label{Pabcdnor}
P_{ABCD}^{\left({\it singlet}\right)} = {1\over 16}P C_{[AB} C_{CD]}=
{\sqrt{3}\over 4}C_{[AB}
C_{CD]}\, d \Phi\, .
\end{equation}
This choice agrees with the normalization of the scalar fields existing
in the current literature.
Let us now consider the gravitino equation (\ref{psisusy}).
Computing the spin connention $\omega^a_{~b}$ from
equation (\ref{ansaetze}), we find
\begin{eqnarray}
\omega^{0i}&=&{d{\cal U}\over dr}{x^i\over r}
e^{{\cal U}\left(r\right)}V^0\nonumber\\
\omega^{ij}&=&2{d{\cal U}\over dr}\, {x_k\over r}\,
 \eta^{k[i}\, V^{j]} \, e^{\cal U}
\end{eqnarray}
where $V^0=e^{\cal U} \, dt$, $V^i = e^{-{\cal U}} \, dx^i$.
Setting $\epsilon_A=e^{f\left(r\right)}\zeta_A$, where $\zeta_A$ is
a constant chiral spinor, we get
\begin{eqnarray}
\label{deltapsi}
&&\left\{{df\over dr}{x^i\over r}e^{f+{\cal U}}\delta_A^BV^i+
\Omega_{A,\alpha}^{~B}\partial_i\Phi^{\alpha}e^fV^i\right.
\nonumber\\
&&\left.-{1\over 4}\left(2{d{\cal U}\over dr}{x^i\over r}e^{\cal U}e^f
\left(\gamma^0\gamma^iV^0+\gamma^{ij}V_j\right)
\right)\delta_A^B \, + \,
\delta_A^B \,T^-_{ab}\gamma^{ab}\gamma^c\gamma^0 V_c\right\}\zeta_B=0
\end{eqnarray}
where we have used eqs.(\ref{psisusy}),(\ref{delepsi}), (\ref{usp8dec}).
This equation has two sectors; setting  to zero the coefficient
of $V^0$ or of $V^i\gamma^{ij}$ and tracing over the $A,B$ indices we find
two identical equations, namely:
\begin{equation}
{d{\cal U}\over dr}=
-{1\over 8} q(r)^{\Lambda\Sigma}{e^{\cal U}\over r^2}
C^{AB}{\rm Im}{\cal N }_{\Lambda \Sigma,\Gamma\Delta} f^{\Gamma\Delta}_{~~AB}.
\label{dudr}
\end{equation}
Instead, if we set  to zero the coefficient of $V^i$, we find
a differential equation for the function $f\left(r\right)$, which
is uninteresting for our purposes.
Comparing now equations (\ref{dphidr}) and (\ref{dudr}) we immediately find
\begin{equation}
\label{phiu}
\Phi= 2 \sqrt{3} \, {\cal U}
\end{equation}

\subsection{Explicit computation of the Killing equations and of the reduced
Lagrangian in the $1/2$ case}

In order to compute the l.h.s. of eq.s
(\ref{finale}), (\ref{dudr}) and the lagrangian of the 1/2 model,
we need the explicit
form of the coset representative $\IL$ given in
equation (\ref{Lcoset}). This will also enable us to compute explicitly the
r.h.s. of equations (\ref{dphidr}), (\ref{dudr}).
In the present case the esplicit form of $\IL$ can be retrieved by
exponentiating the $Usp\left(8\right)$ singlet generator.
As stated in equation (\ref{L-1dL}),
the scalar vielbein in the $Usp\left(28,28\right)$ basis is given by the off
diagonal block elements of $\IL^{-1}d\IL$, namely
\begin{equation}
\label{Pvielbein}
\IP =
\left(
\begin{array}{cc}
0&{\bar P}_{ABCD}\\
P_{ABCD}&0
\end{array}
\right).
\end{equation}
>From equation (\ref{Pabcdnor}), we see that the $Usp\left(8\right)$
singlet corresponds to the generator
\begin{equation}
\label{kappa}
\IK ={\sqrt{3}\over 4}\left( \begin{array}{c|c}
                               0 & C^{[AB} C^{HL]}\\
                               \hline\\
                               C_{[CD}C_{RS]} & 0
                           \end{array}
                \right)
\end{equation}
and therefore, in order to construct the coset representative of the $O\left(1,1\right)$
subgroup of $E_{7(7)}$, we need only to exponentiate $\Phi\IK$.
Note that $\IK$ is a $Usp\left(8\right)$ singlet in the $\bf 70$
representation of $SU\left(8\right)$, but it acts non-trivially in the
$\bf 28$ representation of the quantized charges $\left(e_{AB}, g^{AB}\right)$.
It follows that the various powers of $\IK$ are proportional to the
projection operators onto the irreducible $Usp(8)$
representations $\bf 1$ and $\bf 27$ of the
charges:
\begin{equation}
\label{P1}
\IP_1=\frac{1}{8} C^{AB}C_{RS}
\end{equation}
\begin{equation}
\label{P27}
\IP_{27}=(\delta_{RS}^{AB}- \frac{1}{8} C^{AB}C_{RS}).
\end{equation}
Straightforward exponentiation gives
\begin{eqnarray}
\exp(\Phi\IK)&=&\cosh\left({1\over 2\sqrt{3}}\Phi\right)\IP_{27}+{3\over 2}
\sinh\left({1\over 2\sqrt{3}}\Phi\right)\IP_{27}\IK\IP_{27}+\\
&&+\cosh\left({\sqrt{3}\over 2}\Phi\right)\IP_{1}+{1\over 2}
\sinh\left({\sqrt{3}\over 2}\Phi\right)\IP_{1}\IK\IP_{1}
\end{eqnarray}
Since we are interested only in the singlet subspace
\begin{equation}
\label{P1proj}
\IP_1\exp[\Phi\IK]\IP_1 = \cosh({\sqrt{3}\over 2}\Phi)\IP_1 +
\frac{1}{2} \sinh({\sqrt{3}\over 2}\Phi)
\IP_1\IK\IP_1
\end{equation}
\begin{equation}
\label{Lsinglet}
\IL_{singlet}= \frac{1}{8}\left( \begin{array}{c|c}
                \cosh ({\sqrt{3}\over 2}\Phi) C^{AB}C_{CD} &
                \sinh ({\sqrt{3}\over 2}\Phi) C^{AB} C^{FG}\\
                \hline\\
                 \sinh({\sqrt{3}\over 2}\Phi) C_{CD}C_{LM} &
                 \cosh({\sqrt{3}\over 2}\Phi) C_{CM}C^{FG}
                \end{array}
                \right).
\end{equation}
Comparing (\ref{Lsinglet}) with  the equation (\ref{Lcoset}),
we find \footnote{Note that
we are we are writing the coset matrix with the same couples of
indices $AB, CD, \dots$ without distinction between the couples
$\Lambda\Sigma$ and $AB$ as was done in sect.(\ref{41})}:
\begin{equation}
\label{f}
f=\frac{1}{8 \sqrt{2}} e^{{\sqrt{3}\over 2}\Phi} C^{AB}C_{CD}
\end{equation}
\begin{equation}
\label{h}
h=- \mbox{\rm i} \, \frac{1}{8 \sqrt{2}}\, e^{- {\sqrt{3}\over 2}\Phi} C_{AB}C_{CD}
\end{equation}
and hence, using ${\cal N}=hf^{-1}$, we find
\begin{equation}
\label{Nmatrix}
{\cal N}_{AB \, CD} = - \, \mbox{\rm i} \, {1\over 8}\, e^{-\sqrt{3}\Phi} C_{AB}C_{CD}
\end{equation}
so that we can compute the r.h.s. of (\ref{dphidr}), (\ref{dudr}).
Using the relation (\ref{phiu})  we find a single equation for the unknown functions
${\cal U}\left(r\right)$, $q\left(r\right)=C_{\Lambda\Sigma}q^{\Lambda\Sigma}\left(r\right)$
\begin{equation}
\label{dudr1}
\frac{d{\cal U}}{dr}={1\over 8\sqrt{2}}{q\left(r\right)\over r^2}\exp\left(-2{\cal U}\right)
\end{equation}
At this point to solve the problem completely  we have to consider
also the second order field equation obtained from the lagrangian.
The bosonic supersymmetric
lagrangian of the $1/2$ preserving supersymmetry case is obtained
from equation (\ref{lN=8}) by substituting the values of $P_{ABCD}$
and ${\cal N}_{\Lambda\Sigma|\Gamma\Delta}$ given in equations
(\ref{Pabcdnor}) and (\ref{defN}) into equation (\ref{lN=8}).
We find
\begin{equation}
\label{lbose1/2}
{\cal L} = 2R- e^{- \sqrt{3} \Phi} F_{\mu \nu} F^{\mu \nu} +
\frac{1}{2} \partial_\mu \Phi \partial^\mu \Phi
\end{equation}
\subsection{The $1/2$ solution}
The resulting field equations are
\\
{\underline {\sl Einstein equations:}}
\begin{equation}
\label{einstein1/2}
{\cal U}'' + \frac{2}{r} {\cal U}' - ({\cal U})^2 =
{1\over 4} (\Phi ')^2
\end{equation}
{\underline {\sl Maxwell equations:}}
\begin{equation}
\label{maxwell1/2}
\frac{d}{dr}(e^{-\sqrt{3}\Phi} q(r)) =0
\end{equation}
{\underline {\sl Dilaton equation:}}
\begin{equation}
\label{dilaton1/2}
\Phi'' + \frac{2}{r} \Phi' =-e^{-\sqrt{3}\Phi + 2 {\cal U}}
        q(r)^2 \frac{1}{r^4}   \, .
\end{equation}
>From Maxwell equations one immediately finds
\begin{equation}
q\left(r\right)=e^{\sqrt{3}\Phi\left(r\right)}.
\end{equation}
Taking into account (\ref{phiu}), the second order field equation and the
first order Killing spinor equation have the common solution
\begin{eqnarray}
\label{solution1/2}
{\cal U} & = & -\frac{1}{4} \log \, H(x) \nonumber\\
\Phi     & = & - \frac{\sqrt{3}}{2} \log \, H(x)\nonumber\\
q        & = &  H(x)^{- \frac{3}{2}}
\end{eqnarray}
where:
\begin{equation}
H(x) \equiv 1 + \sum_{\ell} \, \frac{e_\ell}{  {\vec x} - {\vec x}^0_\ell  }
\label{armonie}
\end{equation}
is a harmonic function describing  $0$--branes located at ${\vec x}^0_\ell$ for
$\ell=1, 2,\dots$, each brane carrying a charge $e_\ell$.
In particular for a single
$0$--brane we have:
\begin{equation}
H(x) = 1+ \frac{k}{r}
\label{unasola}
\end{equation}
Note that the lagrangian (\ref{lbose1/2}) and the solution (\ref{solution1/2})
agree with the well known case studied in the literature describing a single
scalar field and a single vector field strength with the peculiar value
$a=\sqrt{3}$ as coefficient of $\Phi$ in the exponential in front of the
vector kinetic term. In the notations of \cite{kstellec} an elementary
$p$--brane solution in space--time dimensions $D$ corresponds to a metric of
the form:
\begin{equation}
ds^2 = H(x_\perp)^{-4 \frac{\tilde d}{\Delta \, (D-2)}} \, dx^\mu _{\parallel}
\otimes dx^\nu_{\parallel} \, \eta_{\mu\nu} +
H(x_\perp)^{4 \frac{  d}{\Delta \, (D-2)}} \,
dx^I _{\perp}
\otimes dx^J_{\perp} \, \delta_{IJ}
\end{equation}
where $H(x_\perp)$ is a harmonic function of the transverse
coordinates, $x_\parallel$ are the parallel coordinates $d=p+1$ is
the dimension of the $p$--brane volume, ${\tilde d}=D-3-p$ is the
dimension of a dual magnetic brane and the {\it dimensional reduction
invariant} $\Delta$ is defined as follows:
\begin{equation}
\Delta = a^2 + 2 \, \frac{d \, {\tilde d}}{D-2}
\label{diminv}
\end{equation}
It is a common wisdom that an elementary $p$--brane preserving $1/2$
of the original string or M--theory supersymmetry has:
\begin{equation}
\Delta = 4
\end{equation}
implying for $0$--branes ($p=0$) in four--dimensions ($D=4$):
\begin{equation}
a=\pm \sqrt{3}
\end{equation}
Our derivation completely confirms this result.
What we have shown is that the most
general $BPS$-saturated black hole preserving $1/2$ of the $N=8$
supersymmetry is actually described by the lagrangian
(\ref{lbose1/2}) with the solution given by (\ref{solution1/2}), in the
sense that any other solution with the same property can be obtained
from the present one by an $E_{\left(7\right)7}$ (U-duality) transformation.
\subsection{Comparison with the Dynkin basis formalism}
The computation of the kinetic matrix
${\cal N}_{\Lambda\Sigma,\Gamma\Delta}$ and of the scalar kinetic
term has been explicitly performed  using the structure of the
$Usp\left(8\right)$ singlet in the so called Young basis, that is in
the basis where the generators of the coset are written in terms of
four index antisymmetric tensors. Alternatively, we could have used
the intrinsic Lie algebra basis in the Weyl--Cartan formalism, which
we name the Dynkin basis. Here
the $Usp\left(8\right)$ singlet generator $H$ among the $70$
generators of the
$E_{\left(7,7\right)}\over SU\left(8\right)$ coset is given as a
suitable linear combination of the noncompact Cartan generators
 $H_{\alpha_i}~~i=1,\dots,7$, dual to the simple roots $\alpha_i$. The
explicit form of such a linear combination is as follows:
\begin{equation}
H=a\left({3\over 2}H_{\alpha_1}+2H_{\alpha_2}+{5\over
2}H_{\alpha_3}+3H_{\alpha_4}+{3\over 2}H_{\alpha_5}+2H_{\alpha_6}+H_{\alpha_7}
\right)
\end{equation}
$a$ being a normalization constant. In the Dynkin Basis $H$ is a
diagonal matrix whose elements are given by
\begin{eqnarray}
&&H={\rm diag}\left(\vec{s},-\vec{s}\right),\\
&&\vec{s}=
\left(-{1\over 2},-{1\over 2},{1\over 2},{1\over 2},{1\over 2},{1\over 2},
-{1\over 2},
-{1\over 2},-{1\over 2},-{1\over 2},{1\over 2},{1\over 2},{1\over 2},
{1\over 2},\right.\\
&&\left.{1\over 2},{1\over 2},
-{1\over 2},{1\over 2},{1\over 2},{1\over 2},{1\over 2},{1\over 2},-{3\over 2}
,-{1\over 2},-{1\over 2},
-{1\over 2},-{1\over 2},-{1\over 2}\right)
\end{eqnarray}
In this basis we immediatly get
\begin{equation}
\IL_D=e^{\Phi H}=
\end{equation}
\begin{equation}
\IL_D^{-1}d\IL_D=Hd\Phi
\end{equation}
where the coset representative $\IL_D$ is written in the $Sp\left(56\right)$
real basis:
\begin{equation}
\IL_D=\left(\begin{array}{cc} A & B \\ C & D \end{array}\right)
\end{equation}
The transformation of  $\IL_D$ from the real $Sp\left(56\right)$ basis to the
complex $Usp\left(28,28\right)$ basis
(see eq. (\ref{Lcoset})) is performed by the the Cayley matrix
(see eq.\eqn{caylone} ) and we get:
\begin{eqnarray}
f&=&{1\over\sqrt{2}}\left(A-\mbox{\rm i} B\right)\\
h&=&{1\over\sqrt{2}}\left(C-\mbox{\rm i} D\right)
\end{eqnarray}
where in our case $B=C=0$.
Therefore, in the Dynkin basis, the kinetic vector matrix is
\begin{equation}
\label{NB=C=0}
{\cal N}=hf^{-1}=-\mbox{\rm i} DA^{-1}
\end{equation}
The explicit listing of the simple roots of $E_{7(7)}$, of the weights
of its ${\bf 56}$--dimensional representation and their
identification with the scalar fields of $M$--theory or of type IIA string
theory compactified on a torus was given in \cite{noi3}. We use
those conventions and in particular we refer to tables 1, 2 of such a
paper. Espressing the vector of the electric and magnetic field strenghts in
this basis we find that the singlet (determined through computation
of the Casimir operator) corresponds to a single electric field
strength at the $23^{rd}$ entry plus a magnetic field strength
at the $28+23$ entry of the general $56-$dimensional vector, all the
other entries being zero. Therefore we get
\begin{equation}
{\cal N}_{\Lambda\Sigma,\Gamma\Delta}F^{\Lambda\Sigma}_{\mu\nu}
F^{\Gamma\Delta\mu\nu}=e^{3a\Phi}F^{23}_{\mu\nu}F^{23\mu\nu}.
\end{equation}
In order to agree with our previous normalization we set
$a={1\over\sqrt{3}}$.
In an analogous way we can compute the scalar kinetic term.
The vielbein is defined as
\begin{equation}
P=Tr\left(\IL^{-1}d\IL
H\right)=Tr\left(H^2d\Phi\right)=18a^2d\Phi
\end{equation}
Therefore
\begin{equation}
{\cal L}_{scal}=k^{11}P_{\mu}P^{\mu}=18a^2\partial_{\mu}\Phi\partial^{\mu}\Phi
\end{equation}
where $k^{11}$ is the one dimensional inverse metric in the $\bf 70$
representation. In our case $k_{11}=Tr\left(HH\right)=18a^2$.
This computation enables us to relate the term
${1\over 6}P_{ABCD\mu}P^{ABCD\mu}$ to
the analogue expression computed in  Dynkin basis. Indeed, in full generality we
have:
\begin{equation}
\label{youngdynkin}
P_{ABCD\mu}P^{ABCD\mu}=
\alpha Tr\left(\IL^{-1}d\IL K_i\right)Tr\left(\IL^{-1}d\IL K_j\right)k^{ij}
\end{equation}
where $K_i,~K_j$ are  $E_{\left(7,7\right)}\over SU\left(8\right)$ coset
generators. In our case we find that, in order to obtain the scalar lagrangian
normalized as in equation (\ref{lbose1/2}), $\alpha ={1\over 2}$
 (once we have fixed $a={1\over\sqrt{3}}$).
The  normalization $\alpha ={1\over 2}$
will be confirmed in next section by the same comparison between
Young and Dynkin formalism at the level of the $1/4$ solution.
\section{Detailed study of the $1/4$ case}
\label{seckilling1/4}
Solutions preserving ${1\over 4}$ of $N=8$ supersimmetry have two
pairs of identical skew eigenvalues in the normal frame for the
central charges. In this case the stability subgroup preserving the
normal form is $O\left(5,5\right)$ with normalizer subgroup in
$E_{\left(7,7\right)}$ given by $SL\left(2,\IR\right)\times O\left(1,1\right)$
(see \cite{FGun}),
according to the decomposition
\begin{equation}
E_{\left(7,7\right)}\supset O\left(5,5\right)\times SL\left(2,\IR\right)\times O\left(1,1\right)=G_{stab}\times G_{norm}
\end{equation}
The relevant fields parametrize ${SL\left(2,\IR\right)\over U\left(1\right)}\times O\left(1,1\right)$ while the
surviving charges transform in the representation $\left({\bf 2,2}\right)$
of $SL\left(2,\IR\right)\times O\left(1,1\right)$. The group
$SL\left(2,\IR\right)$ rotates electric into electric and magnetic into magnetic
charges  while $O\left(1,1\right)$ mixes them. $O(1,1)$ is therefore
a true electromagnetic duality group.
\subsection{Killing spinor equations in the $1/4$ case:surviving
fields and charges.}
The holonomy subgroup $SU\left(8\right)$ decomposes in our
case as
\begin{equation}
SU\left(8\right) \rightarrow  Usp\left(4\right)
\times SU\left(4\right) \times  U\left(1\right)
\end{equation}
indeed in this case the killing spinors satisfy \eqn{urcunquart}
 where we recall the
index convention:
\begin{eqnarray}
&&A,B=1\dots 8 ~~~SU(8)~{\rm indices}\nonumber\\
&&a,b=1\dots 4 ~~~Usp(4)~~~{\rm indices} \nonumber\\
&&X,Y=5\dots 8 ~~~SU(4)~~~{\rm indices}
\end{eqnarray}
and $C_{ab}$ is the invariant metric of $Usp\left(4\right)$.
With respect to the holonomy subgroup
\newline $SU\left(4\right)\times Usp\left(4\right)$,
$P_{ABCD}$ and $T_{AB}$ appearing
in the equations (\ref{chisusy}), (\ref{psisusy}) decompose as follows:
\begin{eqnarray}
\label{7028decomp1/4}
{\bf 70}& \stackrel{Usp(4) \, \times \, SU(4)}{\longrightarrow}&
\left({\bf 1},{\bf 1}\right)\oplus\left({\bf 4},{\bf 4}\right)
\oplus\left({\bf 5},{\bf 6}\right)
\oplus\left({\bf 1},{\bf 6}\right)
\oplus\left({\bf\bar{4}},{\bf\bar{4}}\right)
\oplus\left({\bf\bar{1}},{\bf\bar{1}}\right)
\nonumber\\
{\bf 28}&\stackrel{Usp(4) \,
\times \, SU(4)}{\longrightarrow}&\left({\bf 1},{\bf 6}\right)
\oplus\left({\bf 4},{\bf 4}\right)\oplus\left({\bf 5},{\bf 1}\right)\oplus
\left({\bf 1},{\bf 1}\right)
\end{eqnarray}
We decompose eq. (\ref{chisusy}) according to eq. (\ref{7028decomp1/4}).
We obtain:
\begin{eqnarray}
\delta\chi_{XYZ}&=&0 \\
\delta\chi_{aXY}&=&0 \\
\delta \stackrel{\circ}\chi_{abX} &=&
C^{ab}\delta\chi_{abX}= 0  \\
\delta\chi_{abc}&=&C_{[ab}\delta\chi_{c]} = 0.
\end{eqnarray}
\par
>From $\delta\chi_{XYZ}=0$
we  immediately get:
\begin{equation}
P_{XYZa,\alpha}\partial_{\mu}\Phi^{\alpha}=0
\end{equation}
by means of which we recognize that $16$ scalar fields are actually constant
in the solution.
\par
>From the reality condition
of the vielbein $P_{ABCD}$  (equation (\ref{realcondP}) )
we may also conclude
\begin{equation}
\label{16=0}
P_{Xabc}\equiv P_{X[a}C_{bc]}=0
\end{equation}
so that there are $16$ more scalar fields set to constants.
\par
>From $\delta\chi_{aXY}=0$ we find
\begin{eqnarray}
\label{o111/4}
P_{XY,i}\partial_{\mu}\Phi^i\gamma^{\mu}\gamma^0\epsilon_a&=
&T_{XY\mu\nu}\gamma^{\mu\nu}\epsilon_a\\
\label{30=0}
\stackrel{\circ}{P}_{XYab,i}\partial_{\mu}\Phi^i&=&0
\end{eqnarray}
where we have set
\begin{equation}
\label{decompP1/4}
P_{XYab}=\stackrel{\circ}{P}_{XYab}+{1\over 4}C_{ab}P_{XY}
\end{equation}
Note that equation (\ref{30=0}) sets  $30$ extra scalar fields  to constant.
\par
>From $\delta\chi_{Xab}=0$, using (\ref{16=0}), one  finds
that also $T_{Xa}=0$.

Finally, setting
\begin{eqnarray}
P_{abcd}&=&C_{[ab}C_{cd]}P\\
T_{ab}&=&\stackrel{\circ}{T}_{ab}+{1\over 4}C_{ab}T
\end{eqnarray}
the Killing spinor equation
$\delta\chi_{abc}\equiv C_{[ab}\delta\chi_{c]}=0$ yields:
\begin{eqnarray}
\stackrel{\circ}{T}_{ab}&=&0\\
\label{sl21/4}
P_{,i}\partial_{\mu}\Phi^i\gamma^{\mu}\gamma^0-{3\over
16}T_{\mu\nu}\gamma^{\mu\nu}\epsilon_a&=&0
\end{eqnarray}
Performing the gamma matrix algebra and using equation (\ref{gammaalgebra}),
the relevant evolution
equations (\ref{o111/4}), (\ref{sl21/4}) become
\begin{eqnarray}
\label{sl21/4svil}
P_{,i}{d\Phi^i\over dr}&=&\mbox{\rm i}{3\over
8}\left(g+\mbox{\rm i} q\left(r\right)\right)^{\Lambda\Sigma}
{\rm Im }{\cal N }_{ \Lambda \Sigma,\Gamma\Delta}f^{\Gamma\Delta}_{~~AB}
C^{AB}\, {e^{\cal U}\over r^2}\nonumber\\
P_{XY,i}{d\Phi^i\over dr}&=& 2{\rm i }\left(g+\mbox{\rm i} q\left(r\right)
\right)^{\Lambda\Sigma}{\rm Im } {\cal N }_{ \Lambda \Sigma,\Gamma\Delta}f^{\Gamma\Delta}_{~~XY}
\, {e^{\cal U}\over r^2}
\end{eqnarray}
According to our previous discussion, $P_{XY,i}$ is the vielbein of the
coset $O\left(1,6\right)\over SU\left(4\right)$, which can be reduced to
depend on $6$ real fields $\Phi^i$ since, in force of the
$SU\left(8\right)$ pseudo--reality condition (\ref{realcondP}),
 $P_{XY,i}$  satisfies an analogous pseudo--reality condition.
 On the other hand $P_{,i}$ is the vielbein of $SL\left(2,\IR\right)
\over U\left(1\right)$, and it is intrinsically complex. Indeed
the $SU\left(8\right)$ pseudo--reality condition relates
the $SU\left(4\right)$ singlet $P_{XYZW}$ to the $Usp\left(4\right)$
singlet $P_{abcd}$. Hence $P_{,i}$ depends on a complex scalar field.
In conlusion we find that equations (\ref{sl21/4svil}) are evolutions
equations for $8$ real fields, the $6$ on which $P_{XY,i}$ depends
plus the $2$ real fields sitting in $P_{,i}$.
However, according to the discussion given in section \ref{generaldisc},
we expect that only three scalar fields,
parametrizing ${SL\left(2,\IR\right)\over U\left(1\right)}\times O\left(1,1\right)$
should be physically relevant.
To retrieve this number we note that
$O\left(1,1\right)$ is the subgroup of $O\left(1,6\right)$ which
commutes with the stability subgroup $O\left(5,5\right)$, and hence
also with its maximal compact subgroup $Usp\left(4\right)\times Usp\left(4\right)$.
Therefore out of the $6$ fields of
 $O\left(1,6\right)\over O\left(6\right)$ we restrict our
attention to the real field parametrizing $O\left(1,1\right)$,
whose corresponding vielbein is $C^{XY}P_{XY}=P_1d\Phi_1$. Thus
the second of equations (\ref{sl21/4svil}) can be  reduced to the
evolution equation for the single scalar field $\Phi_1$, namely:
\begin{equation}
\label{reduceo11}
P_{1}{d\Phi^1\over dr}= -2q\left(r\right)^{\Lambda\Sigma}
{\rm Im}{\cal N }_{ \Lambda \Sigma,\Gamma\Delta}f^{\Gamma\Delta}_{~~XY}
\, {e^{\cal U}\over r^2}
\end{equation}
In this equation we have set the magnetic charge
$g^{\Lambda\Sigma}=0$ since, as we show explicitly later, the quantity  $
{\rm Im}{\cal N }_{\Lambda \Sigma,\Gamma\Delta}f^{\Gamma\Delta}_{~~ab}$ is
actually real. Hence, since the left hand side of equation
(\ref{reduceo11}) is real, we are forced to set  the
corresponding magnetic charge to zero.
On the other hand, as we now show, inspection of the gravitino
Killing spinor equation, together with the first of equations
(\ref{sl21/4svil}),  further reduces the number of fields to two.
Indeed, let us consider the $\delta\psi_A=0$ Killing spinor equation.
The starting equation is the same as (\ref{deltapsi}), (\ref{tab}).
In the present case, however, the indices $A,B,\dots$ are $SU\left(8\right)$
indices, which have to be decomposed with respect to
$SU\left(4\right)\times Usp\left(4\right)\times U\left(1\right)$.
Then, from $\delta\Psi_X=0$, we obtain
\begin{equation}
\Omega_X^{~a}=0;~~~T_{Xa}=0
\end{equation}
>From $\delta\psi_a=0$ we obtain an equation identical
to (\ref{deltapsi}) with $SU\left(8\right)$ indices
replaced by $SU\left(4\right)$ indices. With the same computations
performed in the $Usp\left(8\right)$ case we obtain the final
equation
\begin{equation}
\label{dudr1/4}
{d{\cal U}\over dr}=
-{1\over 4}q\left(r\right)^{\Lambda\Sigma}{e^{\cal U}\over r^2}
C^{ab}{\rm Im}{\cal N }_{\Lambda \Sigma,\Gamma\Delta}
f^{\Gamma\Delta}_{~~ab}\, ,
\end{equation}
where we have taken into account
that $C^{ab}{\rm Im} {\cal N }_{\Lambda \Sigma,\Gamma\Delta}
f^{\Gamma\Delta}_{~~ab}$
implying the vanishing of the magnetic charge corresponding
to the singlet of
$U\left(1\right)\times SU\left(4\right)\times Usp\left(4\right)$.
Furthermore, since the right hand side of the
equation (\ref{sl21/4svil}) is proportional to the right
hand side of the gravitino equation, it
turns out that the vielbein $P_i$ must also be real.
Let us name $\Phi_2$ the scalar field appearing in left hand side
of the equation (\ref{sl21/4svil}),
and $P_2$ the corresponding vielbein component.
Equation (\ref{sl21/4svil}) can  be rewritten as:
\begin{equation}
\label{sl21/4fin}
P_2{d\Phi_2\over dr}=-{3\over
8} q\left(r\right)^{\Lambda\Sigma}
{\rm Im}{\cal N }_{ \Lambda \Sigma,\Gamma\Delta}f^{\Gamma\Delta}_{~~XY}
\,C^{XY}{e^{\cal U}\over r^2}
\end{equation}
\par
In conclusion, we see that the most general model describing
$BPS$--saturated solutions preserving $1\over 4$ of $N=8$
supersymmetry is given, modulo $E_{\left(7\right)7}$ transformations,
in terms of two scalar fields and two electric charges.
\subsection{Derivation of the $1/4$ reduced lagrangian in Young basis}
\label{case1/4}
Our next step is to write down the lagrangian for this model.
This implies the construction of the coset representative
of ${SL\left(2,\IR\right)\over U\left(1\right)}\times O\left(1,1\right)$
in terms of which the kinetic matrix
of the vector fields and the $\sigma$--model metric of the scalar
fields is constructed.
\par
Once again we begin by considering such a construction in the Young
basis where the field strenghts are labeled as antisymmetric tensors
and the $E_{7(7)}$ generators are written as $Usp(28,28)$ matrices.
\par
The basic steps in order to construct the desired lagrangian consist
of
\begin{enumerate}
\item Embedding of the appropriate $SL(2,\IR) \times O(1,1)$ Lie
algebra in the $Usp_Y(28,28)$ basis for the ${\bf 56}$ representation of
$E_{7(7)}$
\item Performing the explicit exponentiation of the two commuting
Cartan generators of the above algebra
\item Calculating the restriction of
the $\IL$ coset representative to the $4$--dimensional space spanned
by the $Usp(4) \times Usp(4)$ singlet field strenghts and by their
magnetic duals
\item Deriving the restriction of
the matrix ${\cal N}_{\Lambda\Sigma}$ to the above
$4$--dimensional space.
\item Calculating the explicit form of the scalar vielbein
$P^{ABCD}$ and hence of the scalar kinetic terms.
\end{enumerate}
Let us begin with the first issue. To this effect we consider the
following two antisymmetric $8 \times 8$ matrices:
\begin{equation}
\varpi _{AB} = - \varpi_{BA} = \left (
\begin{array}{c|c}
C & 0 \\
\hline
0 & 0 \\
\end{array} \right) \quad ;  \quad \Omega _{AB} = - \Omega _{BA} = \left (
\begin{array}{c|c}
0 & 0 \\
\hline
0 & C \\
\end{array} \right)
\label{matriciotte}
\end{equation}
where each block is $4 \times 4$ and the non vanishing block $C$
satisfies
\footnote{The upper and lower matrices appearing in $\omega_{AB}$ and
in $\Omega_{AB}$ are actually the matrices $C_{ab}$, $a,b=1,\dots,4$
and $C_{XY}$ $X,Y=1,\dots,4$ used in the previous section.}:
\begin{equation}
C^T = - C \quad ; \quad C^2 = -\bfone
\end{equation}
The subgroup $Usp(4) \times Usp(4) \subset SU(8)$ is defined as the
set of unitary unimodular matrices that preserve simultaneously
$\varpi$ and $\Omega$:
\begin{equation}
A \, \in Usp(4) \, \times Usp(4) \subset SU(8) \quad \leftrightarrow
\quad  A^\dagger \varpi A =
 \varpi \quad \mbox{and}  \quad A^\dagger \Omega A = \Omega
\label{Usp42def}
\end{equation}
Obviously any other linear combinations of these two matrices is also
preserved by the same subgroup so that we can also consider:
\begin{equation}
\tau^\pm_{AB} \, \equiv \, \frac{1}{2}\left( \varpi_{AB} \pm
\Omega_{AB} \right) \,  = \, \left (
\begin{array}{c|c}
C & 0 \\
\hline
0 & \pm C \\
\end{array} \right)
\label{taumatte}
\end{equation}
Introducing also the matrices:
\begin{equation}
 \pi_{AB} \, = \, \left (
\begin{array}{c|c}
\bfone & 0 \\
\hline
0 &  0 \\
\end{array} \right)  \quad ; \quad \Pi_{AB} \, = \, \left (
\begin{array}{c|c}
0 & 0 \\
\hline
0 & \bfone\\
\end{array} \right)
\end{equation}
we have the obvious relations:
\begin{equation}
\pi_{AB} \, = \,  -  \varpi_{AC} \, \varpi_{CB} \quad ; \quad
 \Pi_{AB} \, = \,  -  \Omega_{AC} \, \Omega_{CB}
 \end{equation}
 In terms of these matrices we can easily construct the projection
 operators that single out from the ${\bf 28}$ of $SU(8)$ its
 $Usp(4) \times Usp(4)$ irreducible components according to:
 \begin{equation}
{\bf 28} \quad \stackrel{Usp(4) \times Usp(4)}{\Longrightarrow}\quad
{\bf (1,0)} \oplus {\bf (0,1)}  \oplus {\bf (5,0)}  \oplus {\bf (0,5)}
\oplus {\bf (4,4)}
\label{usp4deco}
\end{equation}
These projection operators are matrices mapping antisymmetric
$2$--tensors into antisymmetric $2$--tensors and  read  as
follows:
\begin{eqnarray}
\IP^{(1,0)}_{AB \  RS} & = & \frac{1}{4} \, \varpi_{AB} \,
\varpi_{RS} \nonumber \\
\IP^{(0,1)}_{AB \ RS} & = & \frac{1}{4} \, \Omega_{AB} \,
\Omega_{RS} \nonumber \\
\IP^{(5,0)}_{AB \ RS} & = & \frac{1}{2} \,\left(
\pi_{AR} \, \pi_{BS} \, - \, \pi_{AS} \, \pi_{BR} \right )\, - \,
\frac{1}{4} \, \varpi_{AB} \, \varpi_{RS}\nonumber \\
\IP^{(0,5)}_{AB \ RS} & = & \frac{1}{2} \,\left(
\Pi_{AR} \, \Pi_{BS} \, - \, \Pi_{AS} \, \Pi_{BR} \right )\, - \,
\frac{1}{4} \, \Omega_{AB} \, \Omega_{RS}\nonumber \\
\IP^{(4,4)}_{AB \ RS} & = & \frac{1}{8} \,\left(
\pi_{AR} \, \Pi_{BS} \,+ \,\Pi_{AR} \, \pi_{BS} \,  - \, \pi_{AS} \, \Pi_{BR}
\, - \, \Pi_{AS} \, \pi_{BR} \right)
\label{proiettori}
\end{eqnarray}
We also introduce the following shorthand notations:
\begin{eqnarray}
\ell^{AB}_{RS} & \equiv & \frac{1}{2} \,\left(
\pi_{AR} \, \pi_{BS} \, - \, \pi_{AS} \, \pi_{BR} \right )\,
\nonumber \\
L^{AB}_{RS} & \equiv &\frac{1}{2} \,\left(
\Pi_{AR} \, \Pi_{BS} \, - \, \Pi_{AS} \, \Pi_{BR} \right )\nonumber\\
U^{ABCD} & \equiv & \varpi^{[AB} \, \varpi^{CD]} \, = \,
\frac{1}{3} \left [ \varpi^{AB} \, \varpi^{CD} + \varpi^{AC} \, \varpi^{DB}\, +
\, \varpi^{AD} \, \varpi^{BC} \right] \nonumber \\
W^{ABCD} & \equiv & \Omega^{[AB} \, \Omega^{CD]} \, = \,
\frac{1}{3} \left [ \Omega^{AB} \, \Omega^{CD} + \Omega^{AC} \, \Omega^{DB}
\,+ \,  \Omega^{AD} \, \Omega^{BC} \right] \nonumber \\
Z^{ABCD}& \equiv & \varpi^{[AB} \, \Omega^{CD]} \, = \,
 \frac{1}{6} \Bigl [ \varpi^{AB} \, \Omega^{CD} + \varpi^{AC} \, \Omega^{DB}
\, \varpi^{AD} \, \Omega^{BC} \nonumber \\
&& + \, \Omega^{AB} \, \varpi^{CD} + \Omega^{AC} \, \varpi^{DB} \, +
\, \Omega^{AD} \, \varpi^{BC} \, \Bigr ]
\label{shortbread}
\end{eqnarray}
Then by direct calculation we can verify the following relations:
\begin{eqnarray}
Z_{ABRS} \, Z_{RSUV} & = & \frac{4}{9} \left( \IP^{(1,0)}_{AB \ UV}
\, + \, \IP^{(0,1)}_{AB \ UV} \, + \, \IP^{(4,4)}_{AB \  UV} \right )
\nonumber \\
U_{ABRS} \, U_{RSUV} & = & \frac{4}{9} \, \ell^{AB}_{UV} \nonumber \\
W_{ABRS} \, W_{RSUV} & = & \frac{4}{9} \, L^{AB}_{UV} \nonumber \\
\label{relazie}
\end{eqnarray}
Using the above identities we can write the explicit embedding of the relevant
$SL(2,\IR) \times O(1,1)$ Lie algebra
into the $E_{7(7)}$ Lie algebra, realized in the Young basis, namely
in terms of $Usp(28,28)$ matrices. Abstractly we have:
\begin{equation}
\begin{array}{ccc}
 \mbox{$SL(2,\IR)$ algebra} & \longrightarrow & \cases { \left[ L_+ \, , \, L_-
 \right] = 2 \, L_0 \cr
  \left[ L_0\, , \, L_\pm
 \right] = \pm \, L_\pm \cr } \\
 \null & \null & \null \\
 \mbox{$O(1,1)$ algebra} & \longrightarrow & {\cal C} \\
 \null & \null & \null \\
 \mbox{and they commute} & \null & \left[ {\cal C} \, , \, L_\pm \right] =
 \left[ {\cal C} \, , \, L_0 \right] = 0 \\
\end{array}
\end{equation}
The corresponding $E_{7(7)}$ generators in the ${\bf 56}$ Young basis
representation are:
\begin{eqnarray}
L_0 & = &  \, \left( \begin{array}{c|c}
 0 & \frac{3}{4}\left( U^{ABFG} + W^{ABFG}\right) \\
 \hline
 \frac{3}{4}\left( U_{LMCD} + W_{LMCD}\right) & 0 \\
 \end{array} \right) \nonumber\\
 L_\pm & = &  \, \left( \begin{array}{c|c}
 \pm \frac{\rm i}{2} \left( \ell^{AB}_{CD} - L^{AB}_{CD}\right) &
 \mbox{\rm i}\, \frac{3}{4}\left( U^{ABFG} - W^{ABFG}\right) \\
 \hline
 -\mbox{\rm i}\, \frac{3}{4}\left( U_{LMCD} - W_{LMCD}\right) &
 \mp \frac{\rm i}{2} \left( \ell^{LM}_{FG} - L^{LM}_{FG}\right) \\
 \end{array} \right) \nonumber\\
 {\cal C}& = & \, \left( \begin{array}{c|c}
 0 & \frac{3}{4}Z^{ABFG} \\
 \hline
 \frac{3}{4}Z_{LMCD} & 0 \\
 \end{array} \right)
\end{eqnarray}
The  non--compact Cartan subalgebra of $SL(2, \IR) \times O(1,1)$, spanned by
$L_0 \, , \, {\cal C}$ is a $2$--di\-men\-sio\-nal subalgebra of the full
$E_{7(7)}$ Cartan subalgebra. As such this abelian algebra is also a subalgebra
of the $70$--dimensional solvable Lie algebra $Solv_7$ defined in eq.
\eqn{defsolv7}. The scalar fields associated with $L_0$ and ${\cal C}$
are the two dilatons  parametrizing the reduced bosonic lagrangian we want to construct.
Hence our  programme is to construct the coset representative:
\begin{equation}
\IL( \Phi_1, \Phi_2) \equiv \exp \left[ \Phi_1 {\cal C}+ \Phi_2 L_0 \right]
\label{cosrepdef}
\end{equation}
and consider its restriction to the $4$--dimensional space spanned by
the $Usp(4) \times Usp(4)$ singlets $\varpi_{AB}$ and $\Omega_{AB}$.
Using the definitions \eqn{proiettori} and \eqn{shortbread} we can
easily verify that:
\begin{eqnarray}
\IP^{(1,0)}_{AB \ RS} \, \frac{3}{4} \, Z^{RSUV}
\, \IP^{(1,0)}_{UV \ PQ} \, &=& 0
\nonumber\\
\IP^{(0,1)}_{AB \ RS} \,\frac{3}{4} \, Z^{RSUV}
\, \IP^{(0,1)}_{UV \ PQ} \, &=& 0
\nonumber\\
\IP^{(1,0)}_{AB \ RS} \,\frac{3}{4} \, Z^{RSUV}
\, \IP^{(0,1)}_{UV \ PQ} \, &=&
\frac{1}{8} \varpi_{AB} \, \Omega_{PQ} \nonumber\\
\IP^{(0,1)}_{AB \ RS} \,\frac{3}{4} \, Z^{RSUV}
\, \IP^{(1,0)}_{UV \ PQ} \, &=&
\frac{1}{8} \Omega_{AB} \, \varpi_{PQ} \nonumber\\
\label{Cgenrest1}
\end{eqnarray}
and similarly:
\begin{eqnarray}
\IP^{(1,0)}_{AB \ RS} \, \frac{3}{4} \, \left( U^{RSUV} +
W^{RSUV} \right)
\, \IP^{(1,0)}_{UV \ PQ} \, &=& \frac{1}{2} \IP^{(1,0)}_{AB \ PQ}
\nonumber\\
\IP^{(1,0)}_{AB \ RS} \, \frac{3}{4} \, \left( U^{RSUV} +
W^{RSUV} \right)
\, \IP^{(0,1)}_{UV \ PQ} \, &=& 0
\nonumber\\
\IP^{(0,1)}_{AB \ RS} \, \frac{3}{4} \, \left( U^{RSUV} +
W^{RSUV} \right)
\, \IP^{(0,1)}_{UV \ PQ} \, &=& \frac{1}{2} \IP^{(0,1)}_{AB \ PQ}
\nonumber\\
\IP^{(0,1)}_{AB \ RS} \,\frac{3}{4} \, Z^{RSUV}
\, \IP^{(1,0)}_{UV \vert PQ} \, &=&
0 \nonumber\\
\label{Cgenrest2}
\end{eqnarray}
This means that in the $4$--dimensional space spanned by the
$Usp(4) \times Usp(4)$ singlets, using also the
definition \eqn{taumatte} and the shorthand notation
\begin{equation}
\Phi^\pm = \frac{\Phi_2 \pm \Phi_1}{2}
\label{shortmilk}
\end{equation}
the coset representative can be written as follows
\begin{eqnarray}
 & \exp\left[\Phi_1 {\cal C} + \Phi_2 L_0  \right]  =&\\
 &  \left( \begin{array}{c|c}
 \cosh \Phi^+\, \frac{1}{2} \, \tau^+_{AB}
 \, \tau^+_{CD} + \cosh \Phi^- \, \frac{1}{2} \, \tau^-_{AB}
 \, \tau^-_{CD} & \sinh \Phi^+ \, \frac{1}{2} \, \tau^+_{AB}
 \, \tau^+_{CD} + \sinh \Phi^-\, \frac{1}{2} \, \tau^-_{AB}
 \, \tau^-_{CD} \\
 \hline
 \sinh \Phi^+ \, \frac{1}{2} \, \tau^+_{AB}
 \, \tau^+_{CD} + \sinh \Phi^-\, \frac{1}{2} \, \tau^-_{AB}
 \, \tau^-_{CD}\ & \cosh \Phi^+ \, \frac{1}{2} \, \tau^+_{AB}
 \, \tau^+_{CD} + \cosh \Phi^- \, \frac{1}{2} \, \tau^-_{AB}
 \, \tau^-_{CD}\\
 \end{array} \right) &
 \label{cassettone}
\end{eqnarray}
Starting from eq.\eqn{cassettone} we can easily write down the
matrices $f_{AB \ CD}, h_{AB \ CD}$ and ${\cal N}_{AB \
CD}$. We immediately find:
\begin{eqnarray}
f_{AB \ CD} & = & \frac{1}{\sqrt{2}} \,
\left( \exp[\Phi^+]  \, \frac{1}{2} \, \tau^+_{AB} \, \tau^+_{CD}  \,
+ \,  \exp[\Phi^-]  \, \frac{1}{2} \, \tau^-_{AB} \, \tau^-_{CD}
\right) \nonumber \\
h_{AB \ CD} & = & -\frac{\rm i}{\sqrt{2}} \,
\left( \exp[-\Phi^+]  \, \frac{1}{2} \, \tau^+_{AB} \, \tau^+_{CD}  \,
+ \,  \exp[-\Phi^-]  \, \frac{1}{2} \, \tau^-_{AB} \, \tau^-_{CD}
\right) \nonumber \\
{\cal N}_{AB \ CD} & = & -\frac{\rm i}{4} \,
\left( \exp[-2 \, \Phi^+]  \, \frac{1}{2} \, \tau^+_{AB} \, \tau^+_{CD}  \,
+ \,  \exp[-2\,\Phi^-]  \, \frac{1}{2} \, \tau^-_{AB} \, \tau^-_{CD}
\right)
\label{fhNmat}
\end{eqnarray}
To complete our programme, the last point we have to deal with is the
calculation of the scalar vielbein $P^{ABCD}$. We have:
\begin{small}
\begin{eqnarray}
& \IL^{-1} (\Phi_1, \Phi_2) \, d \, \IL (\Phi_1, \Phi_2)  \, = \, & \nonumber \\
 & \frac{3}{4} \,\left( \begin{array}{c|c}
 0 & d \Phi_1 \,  Z^{ABFG}   \, + \, d\Phi_2 \,
 \left( U^{ABFG} + W^{ABFG} \right) \\
 \hline
 d \Phi_1 \, Z^{LMCD}   \, + \, d\Phi_2 \,
  \left( U^{LMCD} + W^{LMCD} \right)  & 0 \\
 \end{array} \right) & \nonumber \\
 \label{linv1f}
\end{eqnarray}
\end{small}
so that we get:
\begin{equation}
P^{ABCD} \, = \, d \Phi_1 \, \frac{3}{4} \, Z^{ABCD}   \, + \, d\Phi_2 \,
 \frac{3}{4} \, \left( U^{ABCD} + W^{ABCD} \right)
 \label{scalvielb}
\end{equation}
and with a straightforward calculation:
\begin{equation}
 P^{ABCD}_\mu \, P_{ABCD}^\mu \, =
  \frac{3}{2} \, \partial_\mu \Phi_1 \,\partial^\mu \Phi_1  \, + \,
 3 \, \partial_\mu \Phi_2 \, \partial^\mu \Phi_2 \,
\label{paperino}
\end{equation}
Hence recalling the normalizations of the supersymmetric $N=8$
lagrangian \eqn{lN=8}, and introducing the two $Usp(4) \times Usp(4)$
singlet electromagnetic fields:
\begin{equation}
\label{potentials}
A_\mu ^{AB} = \tau^+_{AB} \, \frac{1}{2 \sqrt{2}} \,{\cal A}^{1}_\mu \,  + \,
\tau^-_{AB} \, \frac{1}{2 \sqrt{2}} \,{\cal A}^{2}_\mu \, + \,
\mbox{26 non singlet fields}
\end{equation}
we get the following reduced Lagrangian:
\begin{eqnarray}
{\cal L}^{1/4}_{red}& = & \sqrt{ -g } \, \Bigl [ 2 \, R[g] \, + \,
\frac{1}{4} \,   \partial_\mu \Phi_1 \, \partial^\mu  \Phi_1
\, + \, \frac{1}{2} \,   \partial_\mu \Phi_2 \, \partial^\mu  \Phi_2\nonumber \\
 &&- \,  \exp\left[-\Phi_1 - \Phi_2 \right]
  \left( F^1_{\mu \nu}\right)^2    \,
  - \,  \exp\left[ \Phi_1 -\Phi_2\right] \left( F^1_{\mu \nu}\right)^2
\label{primalag}
\end{eqnarray}
Redefining:
\begin{equation}
\Phi_1 = \sqrt{2} \, h_1 \quad ; \quad \Phi_2 = h_2
\label{redefo}
\end{equation}
we obtain the final standard form for the reduced lagrangian
\begin{eqnarray}
{\cal L}^{1/4}_{red}& = & \sqrt{ -g } \, \Bigl [ 2 \, R[g] \, + \,
\frac{1}{2} \,   \partial_\mu h_1 \, \partial^\mu  h_1
\, + \, \frac{1}{2} \,   \partial_\mu h_2 \, \partial^\mu  h_2\nonumber \\
 &&- \,  \exp\left[-\sqrt{2} \, h_1 - h_2 \right]
  \left( F^1_{\mu \nu}\right)^2    \,
  - \,  \exp\left[\sqrt{2} \, h_1 - h_2 \right]
  \left( F^2_{\mu \nu}\right)^2
\label{eff1/4}
\end{eqnarray}

\subsection{Solution of the reduced field equations}
We can easily solve the field equations for the reduced lagrangian
\eqn{eff1/4}. The Einstein equation is
\begin{equation}
\label{eqeinstein1/4}
{\cal U}''+{2 \over r}{\cal U}'-\left({\cal U}'\right)^2=
{1 \over 4}\, \left(h_1^\prime \right)^2+ {1 \over 4}\left(h_2^\prime \right)^2,
\end{equation}
the Maxwell equations are
\begin{eqnarray}
\sqrt{2} \, h_1^\prime- h_2^\prime &=&-{q_2' \over q_2}\nonumber\\
-\sqrt{2} \, h_1^\prime- h_2^\prime &=&-{q_1' \over q_1},
\label{eqmaxwell1/4}
\end{eqnarray}
the scalar equations are
\begin{eqnarray}
h_1''+{2 \over r}h_1'&=&{1 \over \sqrt{2} r^4}
\left(e^{ \sqrt{2}h_1- h_2+2{\cal U}}\, q_2^2
-e^{-\sqrt{2}h_1- h_2+2{\cal U}}\, q_1^2\right)\nonumber\\
h_2''+{2 \over r}h_2'&=&{1 \over 2 r^4}
\left(e^{ \sqrt{2}h_1- h_2+2{\cal U}}\, q_2^2
+e^{-\sqrt{2}h_1- h_2+2{\cal U}}\, q_1^2\right).
\label{eqscalars1/4}
\end{eqnarray}
The solution of the Maxwell equations is
\begin{eqnarray}
q_1\left(r\right)&=&q_1e^{\sqrt{2} h_1+h_2}\nonumber\\
q_2\left(r\right)&=&q_2e^{-\sqrt{2} h_1+h_2}
\label{solcharges1/4}
\end{eqnarray}
where
\begin{eqnarray}
q_1&\equiv&q_1\left(\infty\right)\nonumber\\
q_2&\equiv&q_2\left(\infty\right)
\end{eqnarray}
\begin{eqnarray}
h_1&=& -{1 \over \sqrt{2} }\, \log \, {{H_1\over H_2}} \nonumber\\
h_2&=&-{1 \over 2}\, \log \, {H_1 H_2}\nonumber\\
{\cal U}&=&-{1 \over 4}\ln{H_1 H_2}
\label{solfields1/4}
\end{eqnarray}
and
\begin{eqnarray}
H_1\left(r\right)&=&
1 + \sum_{\ell} \, \frac{q^1_\ell}
{ \ {\vec x} - {\vec x}^{(1)}_\ell \ }\nonumber\\
H_2\left(r\right)&=&
1 + \sum_{\ell} \, \frac{q^2_\ell}{ \ {\vec x} - {\vec x}^{(2)}_\ell \ }
\end{eqnarray}
are a pair of harmonic functions.
\subsection{Derivation of the $1/4$ reduced lagrangian in Dynkin basis}
The $1/4$ reduced lagrangian can be alternatively computed in the Dynkin basis.
According to the previous group
theoretical analysis, the lagrangian can depend only on the scalars
that parametrize the normalizer
${SL\left(2,\IR\right)\over U\left(1\right)}\times O\left(1,1\right)$.
\begin{figure}[ht]
\centerline{\epsfig{figure=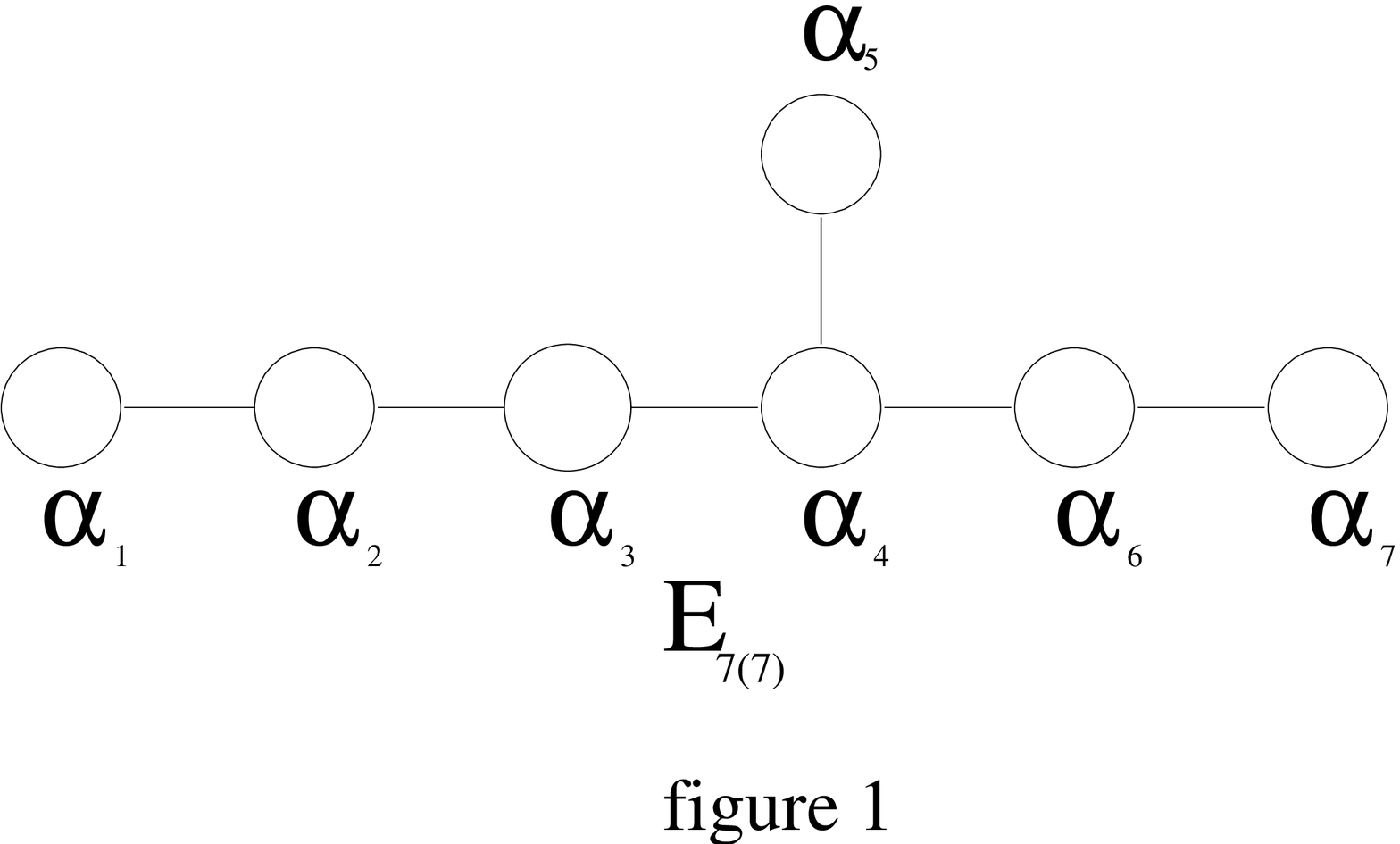,height=5cm,width=10cm,angle=0}}
\end{figure}
Figures $1$ and $2$ show
graphycally the splitting $E_{\left(7\right)7}~\rightarrow~O\left(5,5\right)
\times SL\left(2,\IR\right)\times O\left(1,1\right)$ dictated by the
SLA decomposition
(\ref{stdecomp1a}) (see also \cite{noi2}).
\begin{figure}[ht]
\centerline{\epsfig{figure=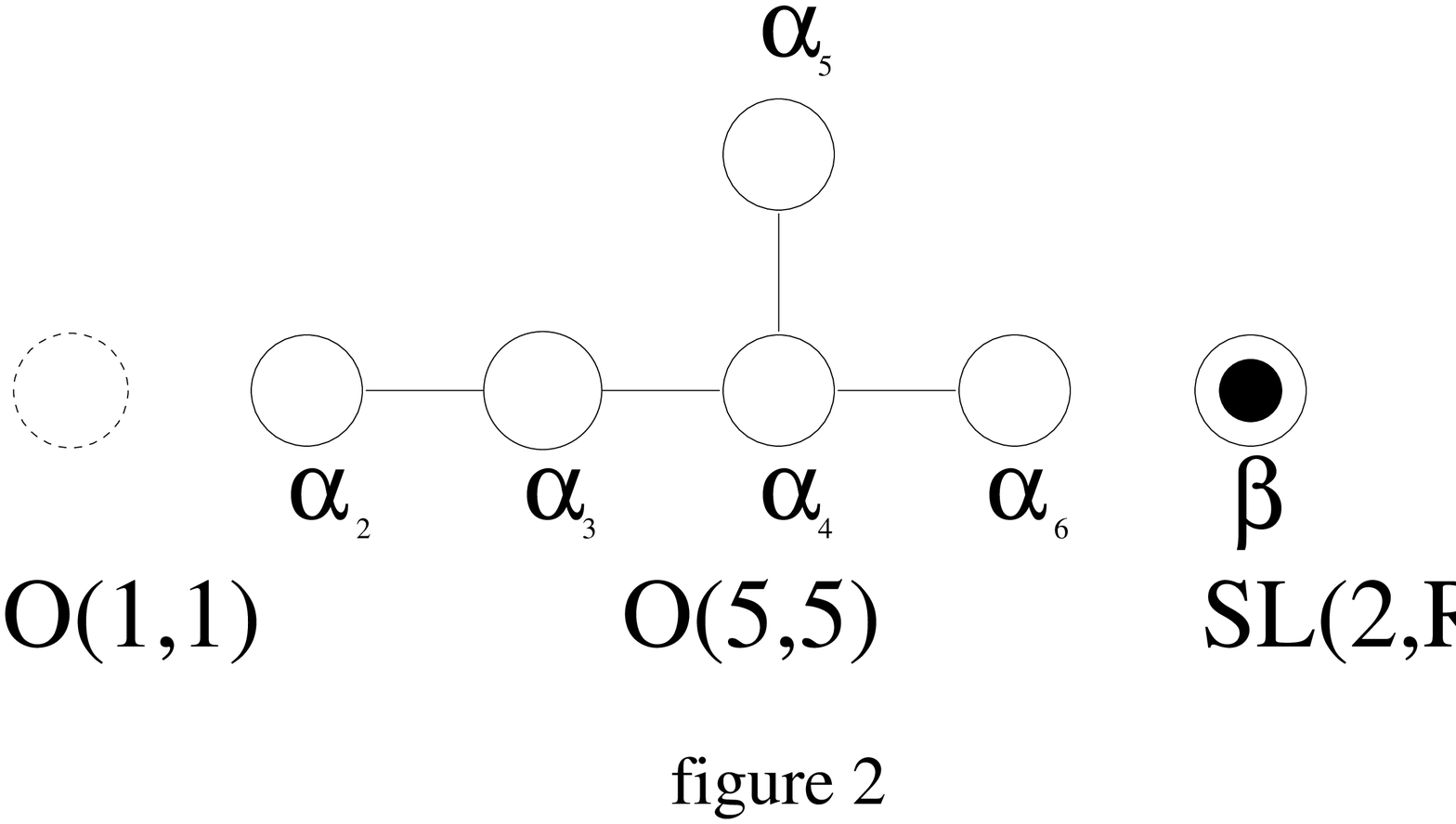,height=5cm,width=10cm,angle=0}}
\end{figure}
The generator of $O\left(1,1\right)$ is the Cartan generator contained in the
$O\left(6,6\right)$ subspace
commuting with  $SL\left(2,\IR\right)$, that is
\begin{equation}
\label{u1gen}
u_1=a\left(H_{\alpha_1}+H_{\alpha_2}+H_{\alpha_3}+H_{\alpha_4}+
    {1 \over 2}H_{\alpha_5}+{1 \over 2}H_{\alpha_6}\right).
\end{equation}
On the other hand, the two solvable ${SL\left(2,\IR\right)\over U\left(1\right)}$ generators are the
Cartan generator and the positive root generator corresponding to the exceptional root $\beta=\alpha_1+2\alpha_2
+3\alpha_3+4\alpha_4+2\alpha_5+3\alpha_6+2\alpha_7$, which commute with $O\left(6,6\right)$ (see \cite{noi2}).
As we have seen from the Killing spinor equations that one of
the two scalars corresponding to these two generators is constant, so we can consider only one of these two generators.

The correct choice of the surviving generator corresponds to the choice of the local $SU\left(8\right)$ gauge fixing.
Indeed, since the $SL\left(2,\IR\right)$ subalgebra commutes with $O\left(5,5\right)\times O\left(1,1\right)$, once the
generators of the noncompact part of $SL\left(2,\IR\right)$ are given, we are free to select which of them is
identified with $H_{\beta}$ and which with $E_{\beta}+E_{-\beta}$. This choice affects our results only when we define
the solvable parametrization of $SL\left(2,\IR\right)\over U\left(1\right)$, which is given in terms of
$H_{\beta}$ and $E_{\beta}$. This is achieved by adding the compact generator $E_{\beta}-E_{-\beta}$ to
$E_{\beta}+E_{-\beta}$. In this way the coset parametrization given
in terms of $\left\{H_{\beta},E_{\beta}\right\}$ becames the solvable
parametrization in terms of $H_{\beta},E_{\beta}$. This procedure
amounts to singling out a particular parametrization of the coset, which implies the $SU\left(8\right)$ gauge fixing.
We choose the Cartan generator
\begin{equation}
\label{u2gen}
u_2=s\left(H_{\alpha_1}+2H_{\alpha_2}+3H_{\alpha_3}+4H_{\alpha_4}+
    2H_{\alpha_5}+3H_{\alpha_6}+2H_{\alpha_7}\right).
\end{equation}
which corresponds to take as scalar field a dilaton instead of
an axion field.

In our Dynkin basis the coset representative is
\begin{equation}
\label{dynkcoset}
\IL_D=e^{\Phi_1u_1}e^{\Phi_2u_2}=\pmatrix{A&B \cr C&D \cr}
\end{equation}
and, since the off-diagonal submatrices $B, C$ are zero, the kinetic matrix (see equation (\ref{NB=C=0})) becomes
\begin{equation}
\label{dynkN}
{\cal N}=-\mbox{\rm i} DA^{-1}.
\end{equation}
The previous analysis of the Killing spinor equations shows that there are no magnetic charges and only two quantized
electric charges corresponding to the $O\left(5,5\right)$ singlets (recall that the quantized charges
belong to the $\bf 56$ of $E_{\left(7\right)7}$). Looking at the quadratic
Casimir operator of $O\left(5,5\right)$, we find that the singlets are at the positions $17,23,45,51$ of the
$56$-dimensional vector, two of them being electric and the other two magnetic. However, we know that the
Killing spinor equations are fulfilled only if both the charges are electric.
To select which of these four charges are electric and which
mangnetic we resort to the equation of motion following from
the lagrangian. It turns out that the equation of motion give a
solution with two electric charges only choosing the singlets
corresponding to the positions $17,51$.
Alternatively, if we have choosen the charges at positions $17,23$,
we would have obtained equations of motion admitting configuration
with one electric and one magnetic charge (and not preserving one fourth
supersymmetry).
We now show that the corret choice gives the same lagrangian
previously obtained via the Young method. We set
\begin{equation}
\label{charges}
F=\cases{F_{17}\equiv F_1\cr F_{51}\equiv F_2\cr {\rm others~zero}\cr}
\end{equation}
and furthermore we fix the normalizations of the scalar fields
setting $a=-{1\over 2},~s={1\over 2}$. Then we find:
\begin{eqnarray}
\label{dynkImNFF}
&&\Im{\cal N}_{\Lambda\Sigma|\Gamma\Delta}F_{\mu\nu}^{~~\Lambda\Sigma}F^{\Gamma\Delta|\mu\nu}=
-\left(e^{-\Phi_1-\Phi_2}F_1^2+e^{\Phi_1-\Phi_2}F_2^2\right)\\
\label{dynkPP}
&&{1 \over 6}P_{ABCD|\mu}P^{ABCD|\mu}={1 \over 12}{\rm Tr}\left(\IL^{-1}d\IL\IK_i\right){\rm Tr}
\left(\IL^{-1}d\IL\IK_j\right)k^{ij}=
\nonumber\\
&&={1\over 4}\partial_{\mu}\Phi_1\partial^{\mu}\Phi_1+{1\over 2}\partial_{\mu}\Phi_2\partial^{\mu}\Phi_2
\end{eqnarray}
reproducing the lagrangian (\ref{eff1/4}), provided the redefinitions (\ref{redefo}) are performed.
In the previous formula $k^{ij}$ is the inverse of the metric of the representation $\bf 70$
 $k_{ij}={\rm Tr}\left(\IK_i\IK_j\right)$. Actually $k_{ij}$ is restricted to the
generators $u_1,~u_2$ and is given by
\begin{equation}
k_{ij}=\pmatrix{3&0\cr 0&6\cr}
\end{equation}

\subsection{Fitting the $1/4$ reduced action into the general framework
of $p$--brane taxonomy}
The reduced action we have found for the $N=8$ black holes with
$1/4$ supersymmmetry
is a particular case of a general action involving $\ell$ scalar
fields $h^i$ ($i=1,\dots,\ell$) and the field strengths of $\ell$ $p+1$--forms
$A^\alpha $  ($\alpha =1,\dots,\ell$), studied by Pope and Lu
\cite{popelu}. The action  reads
\footnote{The normalizations used in this section are the same as we
used in the rest of the paper, except that the field strength
$F_{\mu\nu}$ which in the rest of the paper is defined as
$F=F_{\mu\nu}dx^{\mu}\wedge dx^{\nu}~~~\rightarrow~F_{\mu\nu}={1\over 2}\left(
\partial_{\mu}A_{\nu}-\partial_{\nu}A_{\mu}\right)$ is now defined as
$F_{\mu\nu}=\left(\partial_{\mu}A_{\nu}-\partial_{\nu}A_{\mu}\right)$. The same
observation applies to the $\left(p+2\right)$-forms $F_{\mu_1\cdots\mu_{p+2}}$
which are defined as
$F_{\mu_1\cdots\mu_{p+2}}=\partial_{\mu_1}A_{\mu_2\cdots\mu_{p+2}}+\dots+ {\rm
permutations}$. This explains the presence of the extra factor
$1\over\left(p+2\right)!$ in the $\left(p+2\right)$-form kinetic
lagrangian in (\ref{poplu}).}:
\begin{eqnarray}
S_{PL}&=& \int \, {\cal L}_{PL} \, d^D X \\
{\cal L}_{PL} &=& \sqrt{-g}\left [ 2 \, R[g]+
\frac{1}{2}  \sum_{i=1}^{\ell} \partial_\mu h^i
\partial^\mu  h^i  - \frac{(-1)^{p+1}}{2(p+2)!} \sum_{\alpha
=1}^{\ell} \exp\left[-2{\vec \Lambda}_\alpha \cdot {\vec h}\right]
\left( F^\alpha_{\mu _1 , \dots \mu _{p+2}}\right)^2   \right ]
\label{poplu}
\end{eqnarray}
where $\Lambda_\alpha$ are $\ell$ constant vectors, each  with
$\ell$--components.
In our interpretation $\Lambda_\alpha$ are weights, but what is
crucial in the present discussion is that these weights have a number of
effective components equal to the number of field strenghts in the
game. Our lagrangian \eqn{eff1/4} is of the form \eqn{poplu} with:
\begin{eqnarray}
D&=&4 \quad ; \quad p=0 \quad ; \quad \ell = 2 \nonumber\\
\Lambda_1 &=& \left( 2 \sqrt{2} , 2 \right) \nonumber \\
\Lambda_2 &=& \left( - 2 \sqrt{2} , 2 \right) \nonumber \\
\label{mypeso}
\end{eqnarray}
The field equations derived from the action \eqn{poplu} are the
following ones:
\begin{eqnarray}
&& \mbox{\it Einstein equation:} \\
&&- 2 \, R_{\mu\nu}  = \frac{1}{2} \, \partial_\mu h^i \,\partial_\nu h^i
\, + \, S_{\mu\nu} \quad \label{Einstein}\\
& & \mbox{where}   \nonumber \\
& & S_{\mu\nu}=\frac{(-1)^{p+1}}{2(p+1)!}\, \sum_{\alpha
=1}^{\ell} \exp\left[-2{\vec \Lambda}_\alpha \cdot {\vec h}\right]
\, \left( F^\alpha_{\mu \dots }F^\alpha_{\nu \dots }
-\frac{p+1}{(D-2)(p+2) }\, g_{\mu\nu} \,
F^\alpha_{ \dots }F^\alpha_{ \dots } \right) \\
& & \mbox{\it dilaton equation:} \nonumber\\
& &\frac{1}{\sqrt{-g}} \, \partial_\mu \,\left( \sqrt{-g} \,g^{\mu \nu}
\partial_\nu \, h^i \right)  =  -\frac{(-1)^{p+1}}{2(p+2)!}
\sum_{\alpha =1}^{\ell}\left(-2\Lambda_{\alpha}^i\right)\,
\exp\left[-2{\vec \Lambda}_\alpha \cdot {\vec h}\right] \,
 \left( F^\alpha_{\mu _1 , \dots \mu _{p+2}}\right)^2
 \label{dilaton}\\
 && \mbox{\it Maxwell equation:} \\
 && \partial_\mu \left(\sqrt{-g} \,
 \exp\left[-2{\vec \Lambda}_\alpha \cdot {\vec h}\right] \,
 F^{\alpha \, \mu \nu_1 \dots \nu_{p+1} }\right) = 0
 \label{Maxwell}
\end{eqnarray}

Writing for the metric the usual $p$--brane ansatz:
\begin{equation}
ds^2  =  \exp[2 A(r)] \, dx^a \otimes dx^b \eta_{ab} - \exp[2B(r)] \,
dy^n \, \otimes \, dy^m \, \delta_{nm} \label{metpope} \
\end{equation}
where $x^a$ are the coordinates on the world--volume ($a=0,1,\dots,p$),
$y^n$ are the transverse coordinates ($n=p+1,\dots,D$) and
 $A(y)$,$B(y)$  are two suitable function depending only of the transverse coordinates,
the crucial observation made by Pope and Lu is that,
provided a certain algebraic condition on the {\it weight vectors}
$\Lambda_{\alpha}$ is satisfied, then one can write a
general solution of the differential system \eqn{Einstein},\eqn{dilaton},
\eqn{Maxwell} in terms of exactly $\ell$ harmonic functions:
\begin{equation}
H_{\alpha}(y) = 1\, +\, \sum_{m} \, \frac{k _ {\alpha\, m}}
{({\vec y} - {\vec y}_{\alpha  m)} \, ^{D-p-3} }
\label{harmoalph}
\end{equation}
the constants ${\vec y}_{(\alpha \, m)}$ denoting the location of
the $p$--brane singularities and the constants $k _ {(\alpha\,
m)}$ denoting the charges concentrated on such singular loci.
\par The algebraic condition is as follows. Define the $\ell \,
\times \, \ell $ matrix:
\begin{equation}
M_{\alpha \beta} \equiv 4 \, \Lambda_\alpha  \, \cdot \, \Lambda_\beta
\label{matemme}
\end{equation}
and search for a set of $\ell$ sign choices:
\begin{equation}
\varepsilon_\alpha = \cases{ + 1 \cr
\mbox{or}\cr
-1}
\end{equation}
such that the following equation is satisfied:
\begin{equation}
M_{\alpha \beta}= 4 \, \delta_{\alpha \beta} -  2 \,
 \frac{(p+1) \, (D-p-3)} {D-2} \, \varepsilon_\alpha  \,
 \varepsilon_\beta
 \label{popcondi}
\end{equation}
If a solution to eq.\eqn{popcondi} exists, then in terms of the
corresponding $\varepsilon_\alpha $ and of the $\ell$ harmonic
functions \eqn{harmoalph} the complete BPS solution of the equations of motion
can be written in a universal form. Setting:
\begin{equation}
\varphi_\alpha \equiv -2 \, \Lambda_\alpha \, \cdot \, h
\end{equation}
we have
\begin{eqnarray}
& & \mbox{{\it metric:}} \nonumber \\
 A &=& -\frac{D-p-3}{p+1} \, B \, = \,
 - \frac{1}{2}\, \frac{D-p-3}{D-2} \, \sum_{\alpha =1}^{\ell} \,
\log \, H_\alpha \label{metfie}\\
& & \mbox{\it dilatons:} \nonumber \\
 \varepsilon_\alpha  \, \varphi_\alpha  &=&\log\left[
{H_\alpha ^2}\right]- \, {\frac{(p+1)(D-p-3)}{D-2}} \,
\log\left[ {\prod_{\gamma}^{\ell} H_\gamma
}\right]
\label{difie} \\
& & \mbox{\it Maxwell fields}: \nonumber \\
F^{\alpha} &=& \frac{(-)^{p+1}}{2}\, \left( 1 + \varepsilon_\alpha
\right) \, dx^{a_1} \, \wedge \, \dots \,\wedge dx^{a_{p+1}}
\epsilon_{a_1,\dots\, a_{p+1}} \, dy^m \, \frac{\partial}{\partial y^m}
\, H^{-\varepsilon_\alpha }_\alpha  \nonumber\\
&& + \frac{1}{2}\, \left( 1 - \varepsilon_\alpha \right) \,
dy^{m_1} \, \wedge \, \dots \, dy^{m_{p+2}} \, \epsilon_{m_1\, \dots
\, m_{p+2} \, n} \, \frac{\partial}{\partial y^n} \, H^{-\varepsilon_\alpha }_\alpha
\label{maxwfie}
\end{eqnarray}
As one can see the choice of the $\varepsilon_\alpha$ sign decides
whether the field strength $F^\alpha$ is electric or magnetic. In
dimensions $D$ and for a generic value of $p$ only the electric possibility is
allowed so that in order to get a solution the $\varepsilon_\alpha$
satisfying eq.\eqn{popcondi} should all be positive. Indeed, in order
for the two addends in eq.\eqn{maxwfie}  to make simultaneous sense
it is necessary that
\begin{equation}
D-p-1 = p+2 +1 \quad \longrightarrow \quad D=2(p+2)
\label{dualcondo}
\end{equation}
which is satisfied only in even dimensions and by
$\frac{D-4}{2}$--branes. In this special case which is precisely that
relevant to us, since we have $D=4$, $p=0$, there are  solutions with
both positive and negative $\varepsilon_\alpha$. These are
generalized dyons involving both electric and magnetic charges.
\par
If we apply these general formulae to the case of the lagrangian
\eqn{eff1/4} we can immediately obtain the BPS solution that via
U--duality transformations generates the most general $1/4$ SUSY
preserving black--hole.  Inserting the weight vectors \eqn{mypeso} into
eq.\eqn{matemme} we obtain:
\begin{equation}
M= \left( \matrix{3 & -1 \cr -1& 3}\right)
\label{matmia}
\end{equation}
which by comparison with eq.\eqn{popcondi} leads to the solutions
\begin{equation}
\varepsilon_1 = \varepsilon_2 =1 \quad or \quad \varepsilon_1 = \varepsilon_2 =-1
\label{epsolv}
\end{equation}
So we either have a purely electric or a purely magnetic solution.
If we choose the electric one, specializing  the general formulae
\eqn{metfie},\eqn{difie},\eqn{maxwfie}, we obtain:
\begin{eqnarray}
ds^2 &= &\left(H_1 \,H_2 \right)^{-1/2} \, dt^2 -
\left(H_1 \,H_2 \right)^{1/2} \left( dr^2 + r^2 d\Omega_2 \right)
\nonumber \\
h_1 &=& - \frac{1}{\sqrt{2}}\, \log \frac{H_1}{H_2} \nonumber \\
h_2 &=& - \frac{1}{2} \, \log \left[ H_1 \, H_2 \right]
\nonumber \\
\label{finalsolution}
F^{1,2} &=& -dt \, \wedge \, {\vec {dx}} \,\cdot \,
\frac{\vec \partial}{\partial x} \, \left( H_{1,2} \right)^{-1}
\end{eqnarray}
which is nothing else but the same solution we have already directly
derived from eq.s \eqn{solfields1/4}, \eqn{eqscalars1/4}, \eqn{solcharges1/4}.
Hence we have a perfect fit of the general $1/2$ and $1/4$ supersymmetry
preserving solutions into the general taxonomy of $p$--branes devised
by Pope and Lu.
\subsection{Comparison with the Killing spinor equations}
In this section we show that the Killing spinor equations (\ref{reduceo11}),
(\ref{sl21/4fin}) are identically satisfied by the solution
(\ref{finalsolution}) giving no further restriction on the harmonic
function $H_1,H_2$. Using the formalism developed in section
\ref{case1/4} the equations (\ref{reduceo11}), (\ref{sl21/4fin})
can be combined as follows:
\begin{equation}
\label{16/3}
{16\over 3}P_2{d\Phi_2\over dr}\pm P_1{d\Phi_1\over
dr}=-2q\left(r\right)^{\Lambda\Sigma}\tau^{\left(\pm\right)AB}
{\rm Im}{\cal N }_{ \Lambda \Sigma,\Gamma\Delta}
f^{\Gamma\Delta}_{~~AB}{e^{\cal U}\over r^2}
\end{equation}
where
\begin{eqnarray}
P_2d\Phi_2&=&{3\over 8}C^{ab}C^{cd}P_{abcd}={3\over 8}U^{ABCD}P_{ABCD}={3\over 4}d\Phi_2 \nonumber\\
P_1d\Phi_1&=&C^{XY}C^{ab}P_{XYab}=Z^{ABCD}P_{ABCD}=2d\Phi_1
\end{eqnarray}
furthermore, using equations (\ref{fhNmat}), we have
\begin{equation}
{\rm Im}{\cal N }_{ \Lambda\Sigma,\Gamma\Delta}f^{\Gamma\Delta}_{~~AB}=
-{1\over 2\sqrt{2}}\left(e^{-\Phi_+}\tau^{\left(+\right)\Lambda\Sigma}\tau^{\left(+\right)AB}+
e^{-\Phi_-}\tau^{\left(-\right)\Lambda\Sigma}\tau^{\left(-\right)AB}\right).
\end{equation}
Therefore, the equation (\ref{16/3}) becomes
\begin{equation}
4{d\Phi_2\over dr}\pm 2{d\Phi_1\over dr}=4\sqrt{2}\tau^{\left(\pm\right)}_{\Lambda\Sigma}q^{\Lambda\Sigma}
{e^{{\cal U}-\Phi_{\pm}}\over r^2}
\end{equation}
Using equation (\ref{potentials}) we also have
\begin{eqnarray}
\tau^{\left(+\right)}_
{\Lambda\Sigma}q^{\Lambda\Sigma}&=&{1\over\sqrt{2}}q_1\left(r\right)\nonumber\\
\tau^{\left(-\right)}_
{\Lambda\Sigma}q^{\Lambda\Sigma}&=&{1\over\sqrt{2}}q_2\left(r\right).
\end{eqnarray}
Comparing equation (\ref{finalsolution}) with (\ref{f}),
(\ref{e}), (\ref{smallt}), we obtain
\begin{eqnarray}
q_1&=&-r^2{H'_1\over H_1^2}\sqrt{H_1H_2}\nonumber\\
q_2&=&-r^2{H'_2\over H_2^2}\sqrt{H_1H_2}.
\end{eqnarray}
Using all these informations, a straightforward computation shows that
the Killing spinor equations are identically satisfied.

\section{Conclusions}
This paper extends the analysis of ref. \cite{noi3} to BPS saturated black
holes preserving $1/2$ or $1/4$ of the $N=8$ supersymmetry. Relying on the
SLA approach to the non compact coset space spanned by the scalar fields we
have determined by purely group theoretical tools how many of the scalar
fields are essentially dynamical, in the sense that by an appropriate
$U$-duality rotation all the other can be set to a constant value,
in particular to zero. The same analysis also fixes the number of non--zero
charges and how they transform under the relevant normalizer group of the
stabilizer of the charge in the normal frame.

The Killing spinor equations were also analysed in a group theoretical
fashion and they confirm the predictions of the SLA approach giving in the $1/4$
case the extra information that one of the scalar fields is actually zero.
We were able to solve explicitly the differential equations of the Killing
spinor equations coupled to the Lagrangian of the theory reduced to the relevant
scalar fields and electromagnetic field strengths allowed by our analysis.
These solutions fit nicely in the general framework of ``p-brane taxonomy''
studied in \cite{popelu}.

In conclusion we have now a solution and a thorough group-theoretical
understanding of the $1/2$ and $1/4$ preserving BPS saturated black holes.
The most general solution is obtained by acting on it by a U--duality
transformation on the normal frame configuration.
Note that in ref \cite{noi3}
the same SLA analysis was given for the $1/8$ preserving solution
yielding the result that the STU model is the most general 1/8 model
modulo U--duality transformations. However, the most general solution
could not be derived explicitly, but only one where the axion fields are set
to zero.
It would be nice also in this case to obtain the most general solution. We know
that work is progressing on this issue \cite{trigiamat}.

\par
\vskip 5cm

\vfill
\eject

\begin{thebibliography}{100}

\bibitem{duffrep} M.J. Duff, R.R. Khuri and J.X. Lu, ``{\it String Solitons}'',
Phys.Rept. 259 (1995) 213, hep-th/9412184
\bibitem{kstellec} K. Stelle, ``{\it $BPS$ Branes in Supergravity}'',
 Based on lectures given at the ICTP Summer School in 1996 and 1997.
\bibitem{Polchtasi} For a  review see: J. Polchinski, ``{\it TASI Lectures on
D-Branes}'', hep-th/9611050 and J. Polchinski, S. Chaudhuri and C. Johnson,
``{\it Notes on D-Branes}'', hep-th/9602052
\bibitem{huto2} C. M. Hull and P. K. Townsend, Nucl. Phys. {\bf B 451} (1995)
525, hep-th/9505073 E. Witten, hep-th/9503124, Nucl. Phys. {\bf B443} (1995) 85
\bibitem{malda}  Excellent reviews are: J. M. Maldacena, ``{\it Black Holes in
String Theory}'', Ph. D. Thesis, hep-th/9607235; D. Youm, ``{\it Black Holes
and Solitons in String Theory}'', hep-th/9710046
\bibitem{gensugrabh} References on susy black holes: G. Gibbons, in ``{\it
Unified theories of Elementary Particles. Critical Assessment and
Prospects}'', Proceedings of the Heisemberg Symposium, M\"unchen, West
Germany, 1981,  ed. by P. Breitenlohner and H. P. D\"urr, Lecture Notes in
Physics Vol.  160 (Springer-Verlag, Berlin, 1982); G. W. Gibbons and C. M.
Hull, Phys. lett. {\bf 109B} (1982) 190; G. W. Gibbons, in ``{\it
Supersymmetry, Supergravity and Related Topics}'',  Proceedings of the XVth
GIFT International Physics, Girona, Spain, 1984, ed. by F. del  Aguila, J. de
Azc\'arraga and L. Ib\'a\~nez, (World Scientific, 1995), pag. 147; R. Kallosh,
A. Linde, T. Ortin, A. Peet and A. Van Proeyen, Phys. Rev.  {\bf D46} (1992)
5278; R. Kallosh, T. Ortin and A. Peet, Phys. Rev. {\bf D47} (1993) 5400; R.
Kallosh, Phys. Lett. {\bf B282} (1992) 80; R. Kallosh and A. Peet, Phys. Rev.
{\bf D46} (1992) 5223; A. Sen, Nucl. Phys. {\bf B440} (1995) 421; Phys. Lett.
{\bf B303} (1993) 221;  Mod. Phys. Lett. {\bf A10} (1995) 2081; J. Schwarz and
A. Sen, Phys. Lett. {\bf B312} (1993) 105; M. Cvetic and D. Youm, Phys. Rev.
{\bf D53} (1996) 584; M. Cvetic and A. A. Tseytlin, Phys. Rev. {\bf D53}
(1996) 5619; M. Cvetic and C. M. Hull, Nucl. Phys. {\bf B480} (1996) 296
\bibitem{feka}
S. Ferrara and R. Kallosh, Phys. Rev. {\bf D 54} (1996) 1525, hep/th
9602136
\bibitem{noi3}
L. Andrianopoli, R. D'Auria, S. Ferrara, P. Fr\'e, and M. Trigiante {\it
Nucl.Phys.} {\bf B509} (1998) 463, hep-th/9707087
\bibitem{popelu} C.N. Pope, H.Lu, hep-th/9702086
\bibitem{FGun} S. Ferrara, M. Gunaydin, Int.J.Mod.Phys. A13 (1998) 2075,
hep-th/9708025
\bibitem{noi1}
L. Andrianopoli, R. D'Auria, S. Ferrara, P. Fr\'e,
and M. Trigiante {\it Nucl. Phys.} {\bf B496} (1997) 617,
hep-th/9611014
\bibitem{noi2}
L. Andrianopoli, R. D'Auria, S. Ferrara, P. Fr\'e, R. Minasian
and M. Trigiante {\it Nucl. Phys.} {\bf B493} (1997) 249,
hep-th/9612202
\bibitem{STUkallosh} M. J. Duff, J. T. Liu, J. Rahmfeld, ``{\it Four
Dimensional String/String/String Triality }'', Nucl.Phys. B459 (1996) 125,
hep-th/9508094; R. Kallosh, M. Shmakova, W. K. Wong, "{\it Freezing of Moduli
by N=2 Dyons}", Phys.Rev. D54 (1996) 6284, hep-th/9607077; K. Behrndt, R.
Kallosh, J. Rahmfeld, M. Smachova and W. K. Wond, "{\it STU Black-holes and
string triality}", hep-th/9608059; G. Lopes Cardoso, D. Lust and T. Mohaupt,
``{\it Modular Symmetries of N=2 Black Holes }'', Phys.Lett. B388 (1996) 266,
hep-th/9608099; K. Behrndt, D. Lust, W. A. Sabra, ``{\it Stationary solutions
of N=2 supergravity}'', hep-th/9705169
\bibitem{amicimiei} L. Andrianopoli, R. D'Auria and S. Ferrara, ``{\it
U--Duality and Central Charges in Various Dimensions Revisited}'', Int. Jou.
Mod. Phys. A13 (1998) 431, hep-th/9612105
\bibitem{mylecture} P. Fr\'e, ``{\it Lectures on Special Kahler Geometry and
Electric--Magnetic Duality Rotations}'', Nucl.Phys.Proc.Suppl. 45BC (1996) 59
\bibitem{cre}
E. Cremmer, in ``Supergravity '81'', ed. by S. Ferrara and J.G. Taylor, pag.
313; B. Julia in ``{\it Superspace and Supergravity}'', ed. by S. W. Hawking
and M. Rocek, Cambridge (1981) pag. 331
\bibitem{schw} For  reviews see: J. Schwarz, hep-th/9607201; M. Duff,
hep-th/9608117; A. Sen,  hep-th/9609176
\bibitem{sesch} A. Sen and J. Schwarz, Phys. Lett. {\bf B 312} (1993) 105 and Nucl. Phys.
{\bf B 411} (1994) 35
\bibitem{huto} C.M. Hull and P.K. Townsend, hep-th/9410167,
Nucl. Phys. {\bf B438} (1995) 109
\bibitem{vasch} J. H. Schwarz, ``{\it M--Theory Extension of T--Duality}'',
hep-th/9601077;  C. Vafa, ``{\it Evidence for F--Theory}'', hep-th/9602022
\bibitem{DbranPolch1} J. Polchinski, hep-th/9510017 Phys. Rev. Lett {\bf 75}
(1995) 4724;
\bibitem{myneucha} P. Fr\'e, Talk given at the TMR, EEC contract meeting held
in Neuchatel, september 1997, hep-th/9802045
\bibitem{trigiamat} M. Bertolini, P. Fr\'e, M. Trigiante, work in
progress.
\bibitem{mariotesi} M. Trigiante, Ph.D. Thesis, hep-th/9801144
\bibitem{alex}
D.V. Alekseevskii, Math. USSR Izvestija, {\bf Vol. 9} (1975), No.2
\bibitem{helgason}
S. Helgason, `` {\it Differential Geometry and Symmetric Spaces}'', New York:
Academic Press (1962
\bibitem{recentoine}  P. Claus, M. Derix, R. Kallosh, J. Kumar, P. K. Townsend, A. Van Proeyen, hep-th/9804177

\end{thebibliography}
\end{document}